\documentclass[11pt,titlepage]{article}

\usepackage{setspace}

\usepackage[dvips]{graphicx}
\usepackage{amssymb,amsfonts,amsmath,color}
\usepackage{subfigure}
\usepackage[round,numbers,sort&compress]{natbib}

\usepackage{epsfig, graphicx}
\usepackage{float}
\usepackage{amsmath}
\usepackage{latexsym}
\usepackage{epsfig, graphicx} 
\usepackage{graphics}

\newcommand{\eps}{\varepsilon}

\newtheorem{subn}{\name}

\newcommand{\bsn}[1]{\def\name{#1}\begin{subn}}
\newcommand{\esn}{\end{subn}}
\newtheorem{sub}{\name}[section]
\newcommand{\bs}{\begin{sub}}
\newcommand{\es}{\end{sub}}
\newcommand{\bsl}[1]{\begin{sub}\label{#1}}
\newcommand{\bth}[1]{\def\name{Theorem}\begin{sub}\label{t:#1}}
\newcommand{\blemma}[1]{\def\name{Lemma}\begin{sub}\label{l:#1}}
\newcommand{\bcor}[1]{\def\name{Corollary}\begin{sub}\label{c:#1}}
\newcommand{\bdef}[1]{\def\name{Definition}\begin{sub}\label{d:#1}}
\newcommand{\bprop}[1]{\def\name{Proposition}\begin{sub}\label{p:#1}}

\newcommand{\BA}{\begin{array}}
\newcommand{\EA}{\end{array}}
\newcommand{\BAN}{\renewcommand{\arraystretch}{1.2}
\setlength{\arraycolsep}{2pt}\begin{array}}
\newcommand{\BAV}[2]{\renewcommand{\arraystretch}{#1}
\setlength{\arraycolsep}{#2}\begin{array}}
\newcommand{\BSA}{\begin{subarray}}
\newcommand{\ESA}{\end{subarray}}
\newcommand{\BAL}{\begin{aligned}}
\newcommand{\EAL}{\end{aligned}}
\newcommand{\BALG}{\begin{alignat}}
\newcommand{\EALG}{\end{alignat}}
\newcommand{\BALGN}{\begin{alignat*}}
\newcommand{\EALGN}{\end{alignat*}}
\def\angb<#1>{\langle #1 \rangle}

\newcommand{\noverk}[2]{\left(\begin{array}{c}#1\\#2\end{array}\right)}

\newcommand{\ds}{\displaystyle}
\newcommand{\beq}{\begin{eqnarray}}
\newcommand{\eeq}{\end{eqnarray}}
\newcommand{\beqq}{\begin{eqnarray*}}
\newcommand{\eeqq}{\end{eqnarray*}}

\newcommand{\x}{\mbox{\boldmath$x$}}
\newcommand{\y}{\mbox{\boldmath$y$}}

\def\ds#1{\displaystyle{#1}}

\title{Estimating the synaptic current in a multi-conductance AMPA receptor model}
\author{Adi Taflia,\\ Department of Mathematics,\\ Technion -– Israel Institute of Technology, Haifa, Israel.
\and David Holcman,\thanks{Corresponding author.
E-mail:holcman@biologie.ens.fr}\\ Department Computational
Biology,
\\Ecole Normale Sup\'erieure, 46 rue d'Ulm 75005 Paris, France}
\begin{document}
%

\maketitle
\begin{abstract}
A pre-synaptic neuron releases diffusing neurotransmitters such as
glutamate that activate post-synaptic receptors. The amplitude of
the post-synaptic current, mostly mediated by glutamatergic (AMPARs)
receptors, is a fundamental signal that may generate an action
potential. However, although various simulation results
\cite{kullman,Barbour,Raghavachari} have addressed how synapses control the post-synaptic current,
it is still unclear how this current depends analytically on factors
such as the synaptic cleft geometry, the distribution, the number
and the multi-conductance state of receptors, the geometry of
post-synaptic density (PSD) and the neurotransmitter release
location. To estimate the synaptic current maximal amplitude, we
present a semi-analytical model of glutamate diffusing in the
synaptic cleft. We modeled receptors as multi-conductance channels
and we find that PSD morphological changes can significantly
modulate the synaptic current, which is maximally reliable (the
coefficient of variation is minimal) for an optimal size of the PSD,
that depends on the vesicular release active zone. The existence of
an optimal PSD size is related to nonlinear phenomena such as the
multi-binding cooperativity of the neurotransmitter to the
receptors. We conclude that changes in the PSD geometry can sustain
a form of synaptic plasticity, independent of a change in the number
of receptors.
\end{abstract}



\section*{Introduction}
{S}ynapses are local active micro-contacts underlying direct
neuronal communication. Depending on the brain area and the neuron
types, synapses can vary in size and molecular composition. These
inter-synaptic variations are mediated by hundreds of different
molecules and proteins, participating in the assembly of the stable
but plastic synaptic structure \cite{Sheng,Nicoll1,Nicoll2,Malinow}.
Neurotransmitters such as glutamate molecules, after being released
from vesicles, diffuse in the synaptic cleft, between the pre and
post synaptic terminals (Fig.\ref{figr1}). The post-synaptic
terminal of excitatory synapses contains ionotropic receptors such
as AMPA and NMDA receptors and they may open upon binding with
neurotransmitters. AMPARs are tetrameric assemblies composed of four
different subunits, which can bind to a glutamate molecule
\cite{Cull-candy2006}, but it has been reported that two agonist
molecules at least are required to open a single AMPA channel
\cite{Sakmann}. The amplitude of ionic current is thus proportional
to the number of open receptors and their conductances. The
postsynaptic current measures the efficiency of synaptic
transmission and reports in a complex manner the frequency and
location of released vesicles \cite{Nicoll1}.

It is intriguing that although the number of neurotransmitters
released is of the order of thousands, the number receptors is at
most one hundred.  This difference may serve to compensate the small
receptor patches that should be found by the neurotransmitters
\cite{PNAS}.  To study synaptic transmission, a fundamental step was the
analysis of channels such as AMPARs, expressed in oocytes. By
recording the current using patch clamp and excised patch, the
conductance state properties, related to the open, closed and
desensitized states have been extracted by Markov chain models
\cite{Sakmann,Milstein,Cho}. However, to reduce the complexity of
AMPARs dynamics, Markov chains were kept as minimal as possible,
based on one or two possible binding sites
\cite{Sakmann,Milstein,Cho} with a single conductivity level. Only
recently \cite{Raghavachari}, to study the high variability ($\sim
5-100$ pA) of the synaptic current $I_s$, a four state channel model
was used to interpret the high coefficient of variation, which was
shown to be due to the spatiotemporal correlations of two released
vesicles. When the receptor properties have been sufficiently well
characterized, a second step consisted in integrating these
properties within the synaptic organization to reconstruct the
synaptic function. This step became possible by the use of modeling
and numerical simulations
\cite{Raghavachari,Holmes,Faber,Barbour,Franks,Franks2} allowing to estimate
the role of synaptic geometry on the number of open receptors. More
recently using glutamate diffusion and electrical resistance
properties of the synaptic cleft, it was anticipated that the
synaptic current could be maximal for an optimal cleft height
\cite{Rusakov1}.

We find here by deriving a novel semi-analytical approach that the
synaptic current depends on the vesicular release location, the
number and the biophysical properties of the receptors, the PSD size
and location, and the geometrical characteristics of the synaptic
cleft. To obtain these estimates, we model glutamate as diffusing
molecules and approximate the cleft geometry as a narrow cylinder
\cite{Harris}. But instead of using the classical Markov description
\cite{Sakmann,Milstein,Cho}, our analysis relies on some direct
analysis of AMPA conductances states \cite{Cull-candy2006} and we
account for the four glutamate binding sites per receptor. Our
analysis reveals that given the pre-synaptic active zone size, where
vesicles are released, the coefficient of variation (CV which is the
standard deviation divided by the mean) of the synaptic current is
minimal for a specific PSD size and all other sizes, that may be
induced by plastic changes\cite{Elhers1}, lead to an increasing CV.
However, for a centered active zone, we show that for a fixed
density of receptors, the synaptic current is always a decreasing
function (to zero) of the PSD radius. We further show that a maximal
and reliable current cannot be achieved simultaneously for the same
distribution of synaptic parameters. Finally, we will propose that a
synapse can increase its reliability by restricting the active zone
(AZ) radius, which is a unique and nonlinear function of the PSD
radius. This result should be true for generalized geometry and not
only geodesic disks. Finally, remodeling the PSD, which affects the
synaptic current, can occur in parallel with the classical synaptic
modulation, induced by the direct addition or removal of synaptic
receptors. PSD remodeling can be mediated by geometrical or internal
scaffolding reorganization \cite{El-Husseini,Elhers1}, that can be
transient but much faster than changing permanently the synaptic
receptors. These fast changes can thus affect the detection
threshold of the post-synaptic neuron
\cite{Nicoll,Elias,El-Husseini} and be a source of synaptic
plasticity without changing the number of AMPARs.

\section*{Method: Theoretical model}
\subsection*{Diffusion in the synaptic cleft.} The synaptic current
$I_s$ is mediated by open AMPARs, which can bind from one to four
glutamate molecules. A single AMPAR has several conductance states
which correspond to the combination of the four distinct
conductances, associated to the different GluR subunits
\cite{Stevens,Cull-candy2006}. However, to reduce the complexity
of the analysis, most studies
\cite{Sakmann,Milstein,Cho,Holmes,Barbour,Franks} have modeled
AMPARs dynamics by one or two bound glutamate molecules. The
conductances designated by $\gamma_1,\gamma_2,\gamma_3,\gamma_4$
associated with 1, 2, 3 or 4 bound glutamate molecules. It has
been reported \cite{Gouaux1,Gouaux2} that at least two glutamate
molecules are needed to open an AMPAR, so we set $\gamma_1=0$. The
synaptic current $I_s(t)$ depends on the number of open channels
$(N_2,N_3,N_4)$ bound respectively by 2,3 or 4 glutamate
molecules. In that case,
\beq \label{current} I_s(t)=(\gamma_2
N_2(t)+\gamma_3 N_3(t)+\gamma_4 N_4(t)) \Delta V,
\eeq
where $\Delta V$ is the difference of potential between the intra
and the extracellular medium. We selected from the experimental
results \cite{Cull-candy2006}, obtained for different glutamate
concentration the values $\gamma_2=4 pS$, $\gamma_3=10pS$,
$\gamma_4=13pS$ (for $\Delta V_m=-100 mV$).

\begin{figure}
   \begin{center}
      \includegraphics*[width=3.25in]{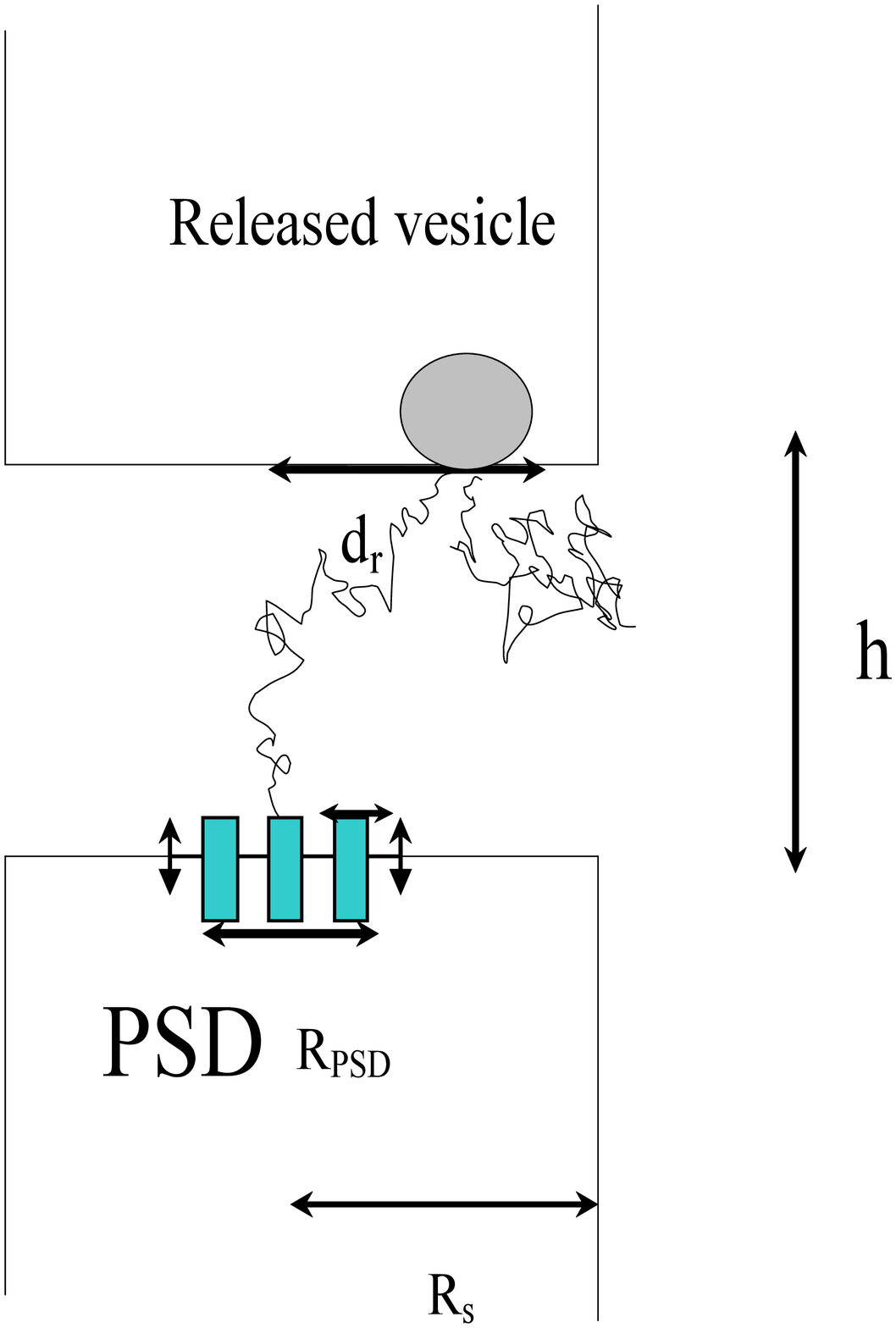}
      \caption{}
      \label{figr1}
   \end{center}
\end{figure}

Our goal here is to estimate the mean and the variance of open
receptors and quantify the peak amplitude of the current $I_{s}$.
After a vesicle fuses with the pre-synaptic membrane at position
$\x_0$, $N_g=3000$ glutamates are released (Fig.\ref{figr1}).
Glutamate molecules diffuse in the cleft and are reflected on the
synaptic membrane, while they are absorbed at the lateral boundary
of the synaptic cleft. AMPARs are uniformly distributed over the
PSD. We consider that only a certain fraction of glutamates hitting
an AMPAR leads to receptor activation, due to a chemical energy
barrier. The synaptic cleft is modeled as a cylinder $\Omega$
(Fig.\ref{figr1}), while there are $N_a$ AMPARs located on the PSD.
When a glutamate molecule hits an AMPAR, it can either be reflected
or it will activated the receptor. We model this behavior using a
homogenized radiative boundary condition over the entire PSD
($\partial
\Omega_{PSD}$) \cite{Zwanzig,Berezhkovskii,Reingruber}. A glutamate
that hits the neuron membrane is reflected except at the lateral
cleft boundary ($\partial \Omega_{Lat}$), where it will not
contribute to activate an AMPAR and this is modeled by an absorbing
boundary condition.
\subsection*{Estimating the number of open receptors.}\label{appendix1}
The probability to find a glutamate molecule at position $\x$ at
time $t$, when it started at position $\x_0$ is given by the density
function $p(\x,t|\x_0)$ that satisfies the equation: \beq
\label{diffuseq} \frac{\partial p(\x,t|\x_0)}{\partial t} &=&D\Delta
p(\x,t|\x_0), \quad
\quad\x \in\Omega, \ t>0 \\
p(\x,0)&=&\delta(\x-\x_0) \nonumber \\
\left.\frac{\partial p(\x,t|\x_0)}{\partial\nu}\right|_{\partial \Omega_r} &=&0,
\, \, \, \, \, \,
\left.p(\x,t|\x_0)\right|_{\partial\Omega_{Lat}}=0 \nonumber \\
& & \nonumber \\
-D\left.\frac{\partial p(\x,t|\x_0)}{\partial\nu}\right|_{
\partial\Omega_{PSD}}& =&-\kappa p(\x,t|\x_0),\nonumber
\eeq
where $D$ is the free glutamate diffusion constant. The partial
absorption constant $\kappa$ accounts for the fraction of AMPARs
inside the PSD and the activation barrier of a glutamate to a GluR
binding site. Using a homogenization procedure
\cite{Szabo,Zwanzig,Berezhkovskii,Reingruber},
we shall derive a new expression in the context of the synaptic
cleft for the partial reflection parameter $\kappa$
\beq \label{eq45ant}
\kappa =\frac{D}{2\pi R^2_{PSD}}\frac{1}{\frac{f(\sigma)}{N_a a}+\frac{D}{\kappa_a 2 \pi
a^2 N_a} },
\eeq
where $f(\sigma)=1-\sigma,$ $\sigma=N_aa^2/R^2_{PSD}$, $a$ is the
radius of the binding site and $R_{PSD}$ is the radius of the PSD
and $\kappa_a$ measures the partial binding of a glutamate molecule
to a single receptor. We have summarize in the appendix all these
steps. To determine the constant $\kappa_a$,  we used the fitted
results obtained from the Markov analysis given in \cite{Milstein}.
In a first approximation neglecting $\kappa_a$, the partial
reflecting constant $\kappa$ is directly proportional to the binding
rate of glutamate molecules and in the appendix, we obtain the
numerical approximation $\kappa_a \approx 1.06$.

To determine the value of $\kappa$ (see appendix \ref{appen2}) we
further need to estimate the effective radius $a$ of a single
receptor. For that purpose, we run some simulations for a typical
synapse of radius $500nm$ with a PSD radius of $300nm$, a height of
$30nm$ and an AZ radius of $150nm$. Using the criteria  that the
synaptic current saturates for four released vesicles ($ \sim
12,000$ glutamate molecules), we obtain that radius $a \approx 1.8
nm$, as shown in figure (\ref{fig:estimate_of_effectin ra}).
Interestingly, the radius $a$ accounts not only for the geometrical
properties of the AMPAR binding site, but also for the underlying
electrostatic interactions. This value $a$ should be compared to the
recent crystal structure dimensions (the AMPAR has a transversal
size of 9nm and the total length is around 18nm) of the
ligand-binding domain reported to be less than 4nm
\cite{AMPA-struct}.

The probability $p(\x_0)$ that a glutamate molecule released at
position $\x_0$, binds a receptor is given by the total flux:
\beq
p(\x_0)=-D\int_0^\infty\int_{\partial\Omega_{PSD}}\frac{\partial
p(\y,t|\x_0)}{\partial n}\,d\y dt= -D \int_{\partial\Omega_{PSD}}
\frac{\partial u(\y|\x_0)}{\partial n}d\y \nonumber
\eeq
where
$u(\x|\x_0)=\int_0^{\infty} p(\x,t|\x_0)dt$
satisfies
\beq
\label{dynb}
 D\Delta u(\x|\x_0)&=&-\delta(\x-\x_0)\quad\mbox{for}\quad
\x\in\Omega\nonumber \\
& &\nonumber\\
\left.\frac{\partial u(\x|\x_0)}{\partial\nu}\right|_{\partial \Omega_r} =0, \, &&
\left.u(\x|\x_0)\right|_{\partial\Omega_{Lat}}=0 \nonumber \\
 & & \nonumber \\
D\left.\frac{\partial u(\x|\x_0)}{\partial\nu}\right|_{
\partial\Omega_{PSD}}& =&-\kappa u(\x|\x_0),\nonumber
  \eeq
In the appendix, we derive an analytical expressions of $u$ and $p$.
For a vesicle releasing its contain at the center of the AZ, the
probability to bind one of the AMPARs is
\beqq
p=\frac{J_{PSD}}{J_{PSD}+J_{Lat}}.
\eeqq
where
\beqq 
 J_{PSD}=2\pi\kappa\int_0^L\frac{1}{ D}K_0(\alpha
r)rdr+2\pi\kappa\int_0^L AI_0(\alpha r)rdr
\eeqq
and
\beqq
J_{Lat}=  -2\pi D h C.
\eeqq
where A and C are two constants that depend on L,  $\kappa$, D, and
h, given in the appendix.
\subsection*{ The mean and variance of the synaptic current
$I_s$.}\label{appendix2}
To compute the mean and the variance of the maximal amplitude of
synaptic current $I_s$, we shall account for two possible sources of
fluctuations: one is due to the number of bound glutamate molecules
and the second to the configuration of bound AMPARs. By
configuration, we mean the distribution of AMPARs bound to 2, 3 and
4 glutamate molecules.

To estimate the first source, we use the probability distribution
$Pr_k(\x_0)$ to have $k$ glutamate molecules bound, when a vesicle
is released at position $\x_0$. This probability follows a binomial
distribution
\beq \label{eqproba}
Pr_k(\x_0) =C^k_{N_g} p(\x_0)^k
(1-p(\x_0))^{N_g-k} \eeq \beq \label{probax0} p(\x_0)=\kappa
\int_{\partial\Omega_{PSD}} u(\x|\x_0) dS(\x),
\eeq
where $N_g$ is the total number of released glutamate molecules.
Because AMPARs can bind from zero to four glutamate molecules, the
probability of a given configuration $\vec{n}=(n_4,n_3,n_2,n_1)$ to
have $n_1$ AMPARs bound to one glutamate, $n_2$ AMPARs bound to two
glutamates and so on, when there are $N_a$ AMPA receptors, for
$k\leq min (4N_a,N_g)$ bound glutamate molecules, is given by
\beq\label{combi}
\Pr\{\vec{n}|k\}=\frac{N_a!}{n_4!n_3!n_2!n_1!(N_a-(n_4+n_3+n_2+n_1))!}\frac{1}{F(k,N_a)},
\eeq
where this probability is computed by choosing $n_4$ AMPARs out of
$N_a$, $n_3$ out of $N_a-n_4$ and so on. $F(k,N_a)$ is the number of
possibilities to decompose the integer $k$  on the integer
$4,3,2,1,0$, when there are at most $N_a$ terms: that is
$k=4n_4+3n_3+2n_2+n_1.$ In practice, we compute $F(k,N_a)$
numerically as the $k+1$'s coefficient of the expression
$(1+x+x^2+x^3+x^4)^{N_a}$. The present analysis can be used to
obtain any statistical moments associated to the current and in
particular, the mean and variance
\beq \label{syncurrent}
\langle I_s(\x_0)\rangle&=& \Delta V\sum_{k=1}^{N_g}\sum_{n \in S_k }\vec{n}\cdot\vec{\gamma}\Pr\{\vec{n}|k\} Pr_k(\x_0)= \\
&&\Delta V\sum_{k=1}^{4N_a}\sum_{n \in S_k
}\vec{n}\cdot\vec{\gamma}\Pr\{\vec{n}|k\} Pr_k(\x_0)+\nonumber
\\&&\Delta V N_a\gamma_4 \left(
1-\sum_{k=0}^{4N_a}Pr_k(\x_0)\right)\nonumber\eeq \beq \langle
I^2_s(\x_0)\rangle&=&\Delta V^2\sum_{k=1}^{N_g}\sum_{n \in S_k}(
\vec{n}\cdot\vec{\gamma})^2 \Pr\{\vec{n}|k\}Pr_k(\x_0)-\langle
I_s(\x_0)\rangle^2=
\nonumber\\
&&\sum_{k=1}^{N_a}\sum_{n \in S_k}( \vec{n}\cdot\vec{\gamma}\Delta
V)^2\Pr\{\vec{n}|k\}Pr_k(\x_0)+\nonumber\\&&(N_a\gamma_4 \Delta
V)^2 \left(1-\sum_{k=0}^{4N_a}Pr_k(\x_0)\right)-\langle
I_s(\x_0)\rangle^2\nonumber
\eeq
where $\Delta V$ is the voltage drop,
$\vec{\gamma}=(\gamma_1,\gamma_2,\gamma_3,\gamma_4)$ is the
conductances vector, $S_k$ represents the set of possible
configurations of $\vec{n}=(n_1,n_2,n_3,n_4)$ such that
$4n_4+3n_3+2n_2+n_1=k$. The formulas for the mean and variance are
made of two terms: the first is the sum over all the sites that are
partially bound by glutamate molecules and the probability for such
an event is the product of the probability $Pr_k$ that $k$
glutamates are bound ($k <4N_a$) and the probability
$\Pr\{\vec{n}|k\}$ of a given binding configuration
$k=4n_4+3n_3+2n_2+n_1$. The second term accounts for all bound
AMPARs ($4N_a$) and this happens with the complementary probability
to the first case.

\section*{Numerical simulation results for the synaptic current}
\begin{figure}
   \begin{center}
      \includegraphics*[width=3.25in]{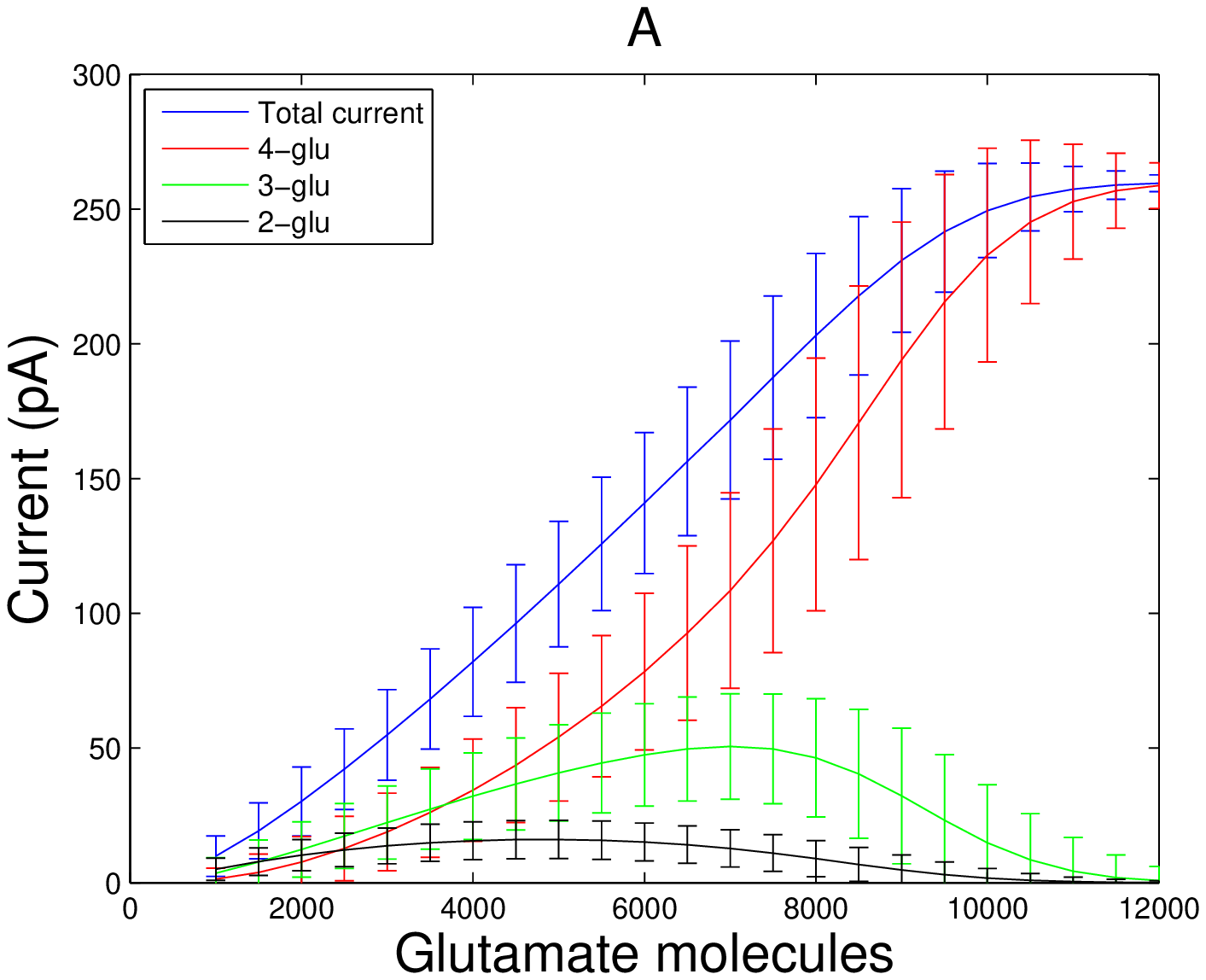}
        \includegraphics*[width=3.25in]{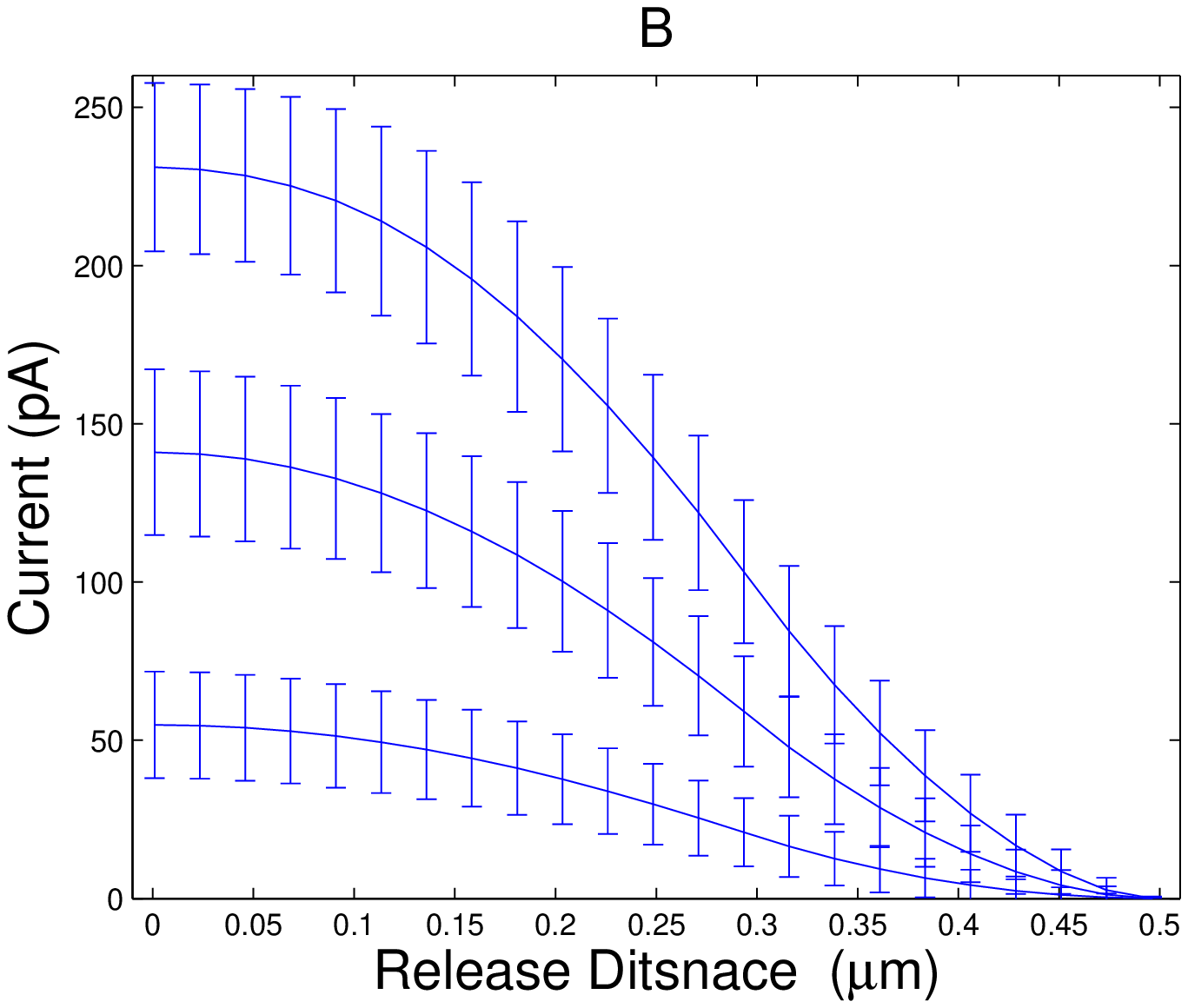}
      \caption{}
      \label{figr2}
   \end{center}
\end{figure}

\begin{figure}
   \begin{center}
      \includegraphics*[width=3.25in]{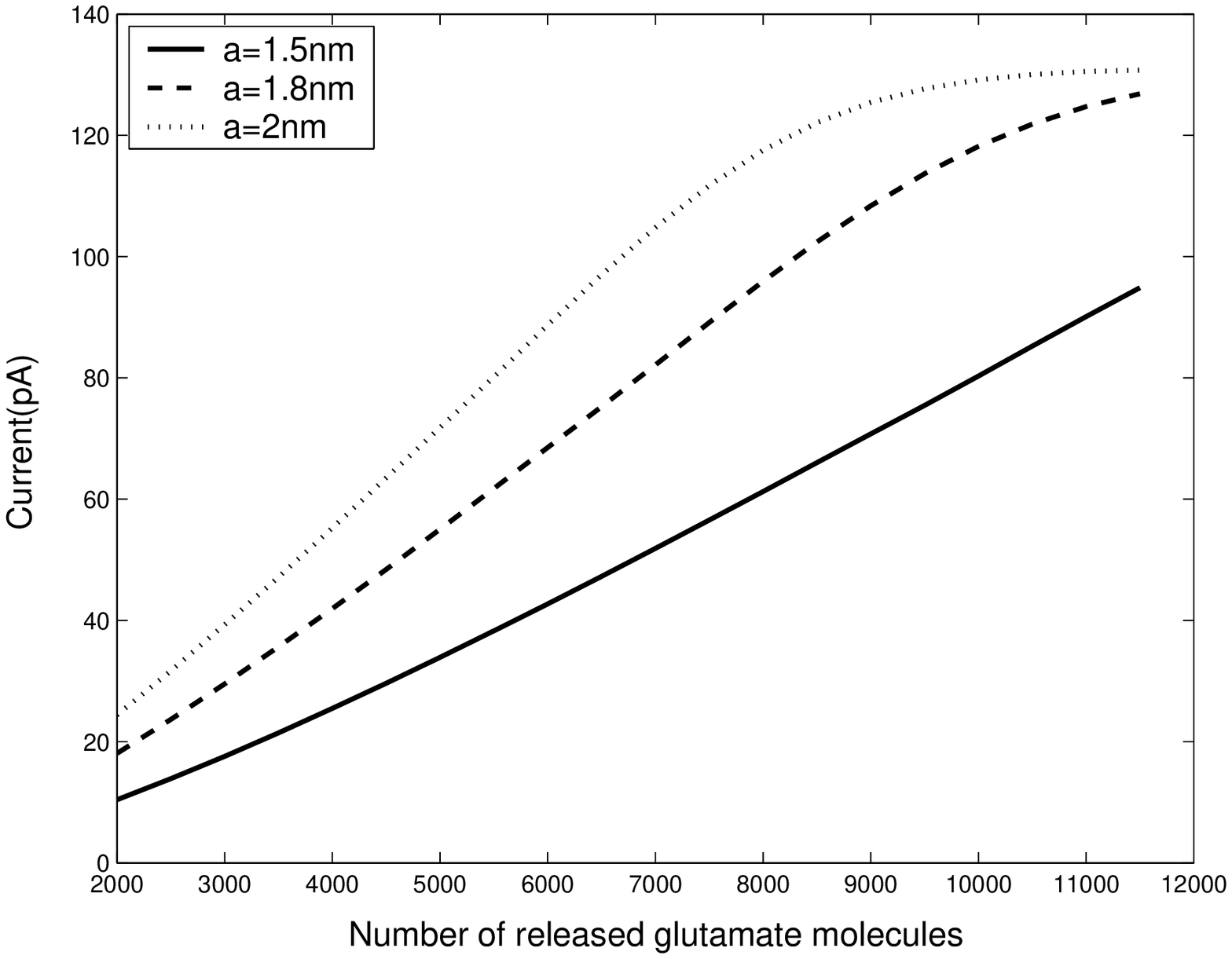}
      \caption{}
      \label{fig:estimate_of_effectin ra}
   \end{center}
\end{figure}

To determine the maximum amplitude of the synaptic current, we use
our semi-analytical method developed above and simulate a single
vesicle ($N_g=3000$) released. We find (Fig.\ref{figr2}A) that
AMPARs are mostly bound to two glutamate molecules, while for
already two vesicles, the dominant contribution comes from receptors
bound to four. To further investigate the influence of the PSD size
on the synaptic current, We plotted the current $I_s$ as a function
of the release site position (Fig.\ref{figr2}B):  when release
occurs outside the region above the PSD, the current drops
drastically due to a fast decay of the binding probability.

\begin{figure}
   \begin{center}
   \begin{tabular}{ll}
      \epsfig{file=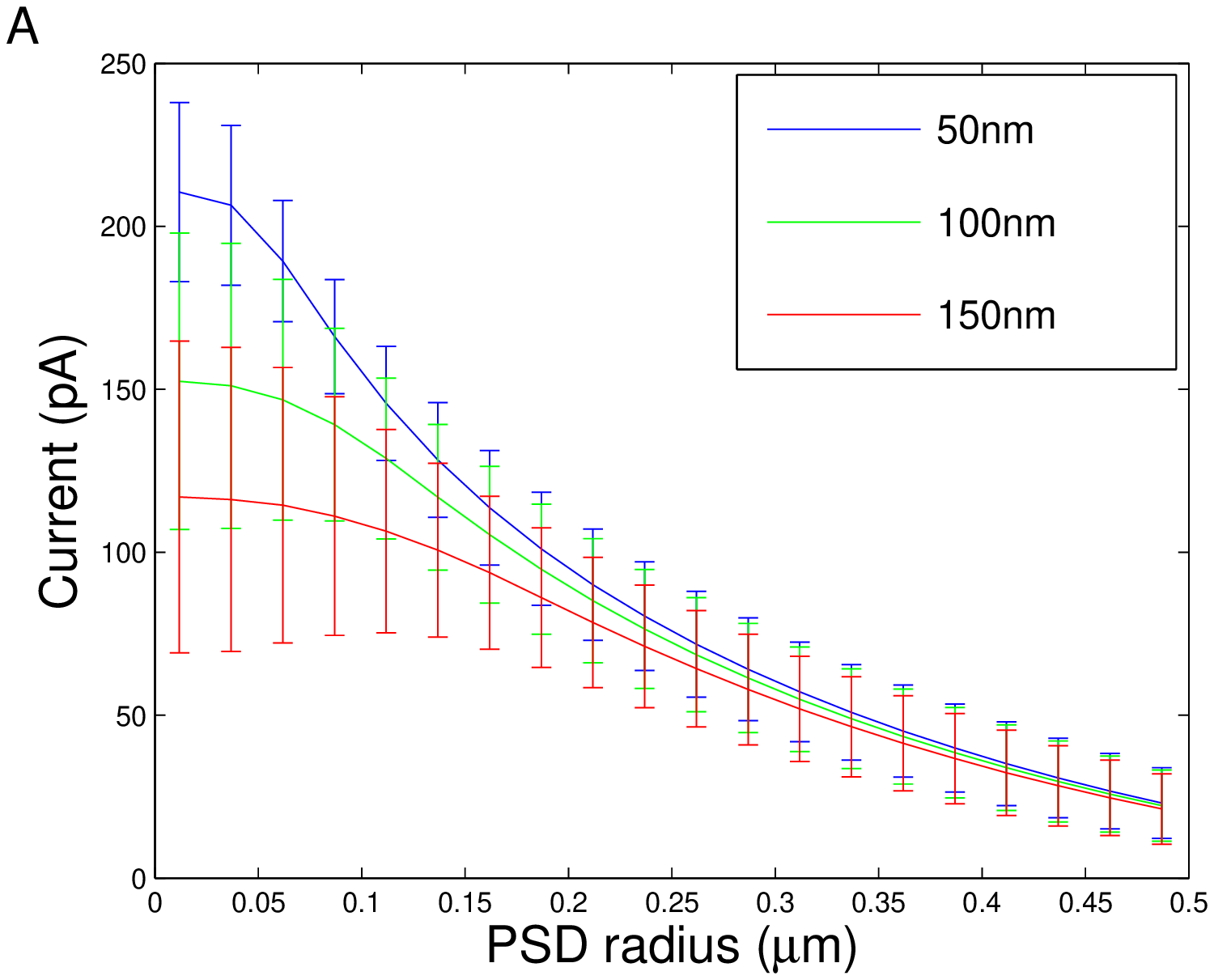,width=3.25in,clip=}&
        \epsfig{file=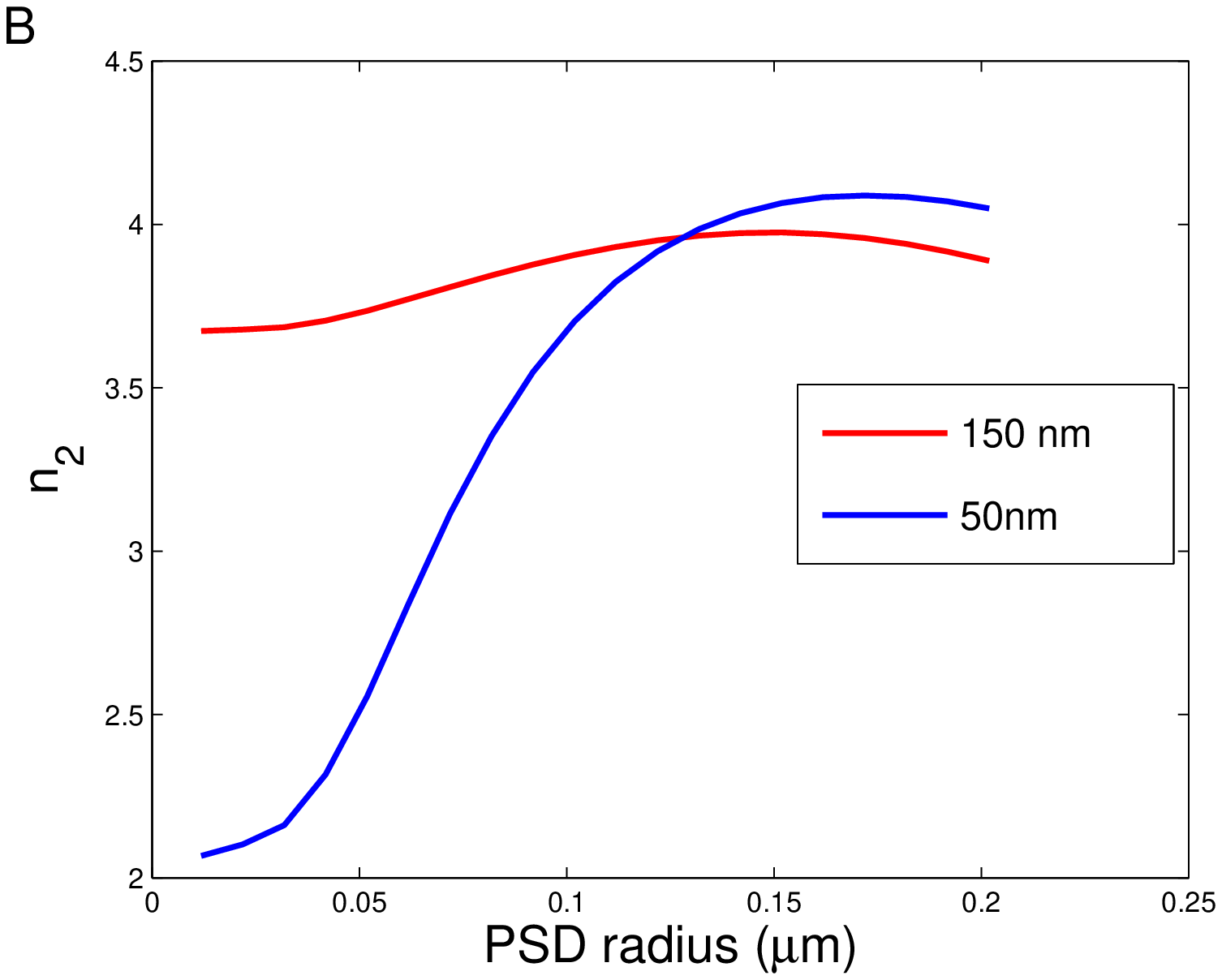,width=3.25in,clip=}\\
            \epsfig{file=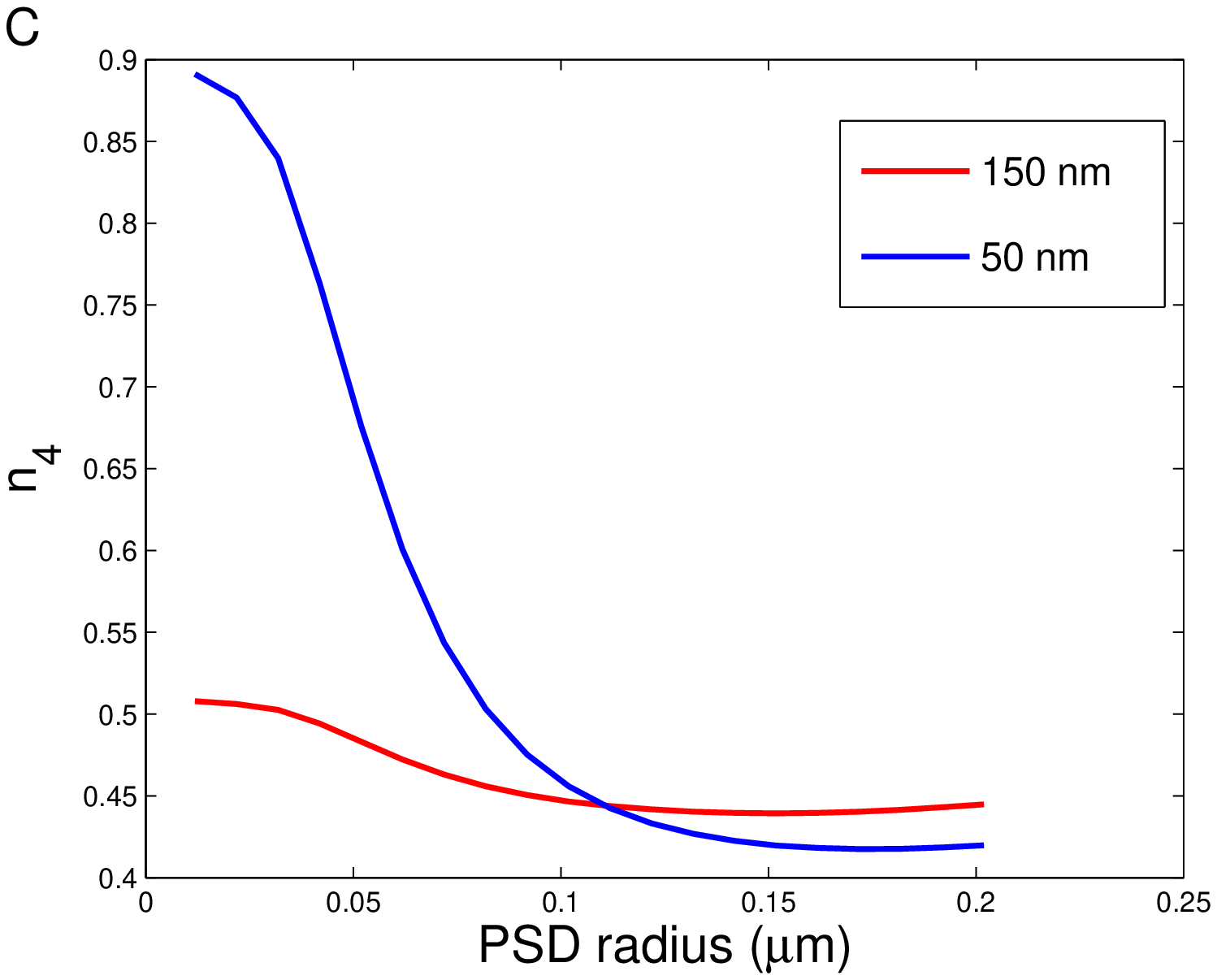,width=3.25in,clip=}&
                \epsfig{file=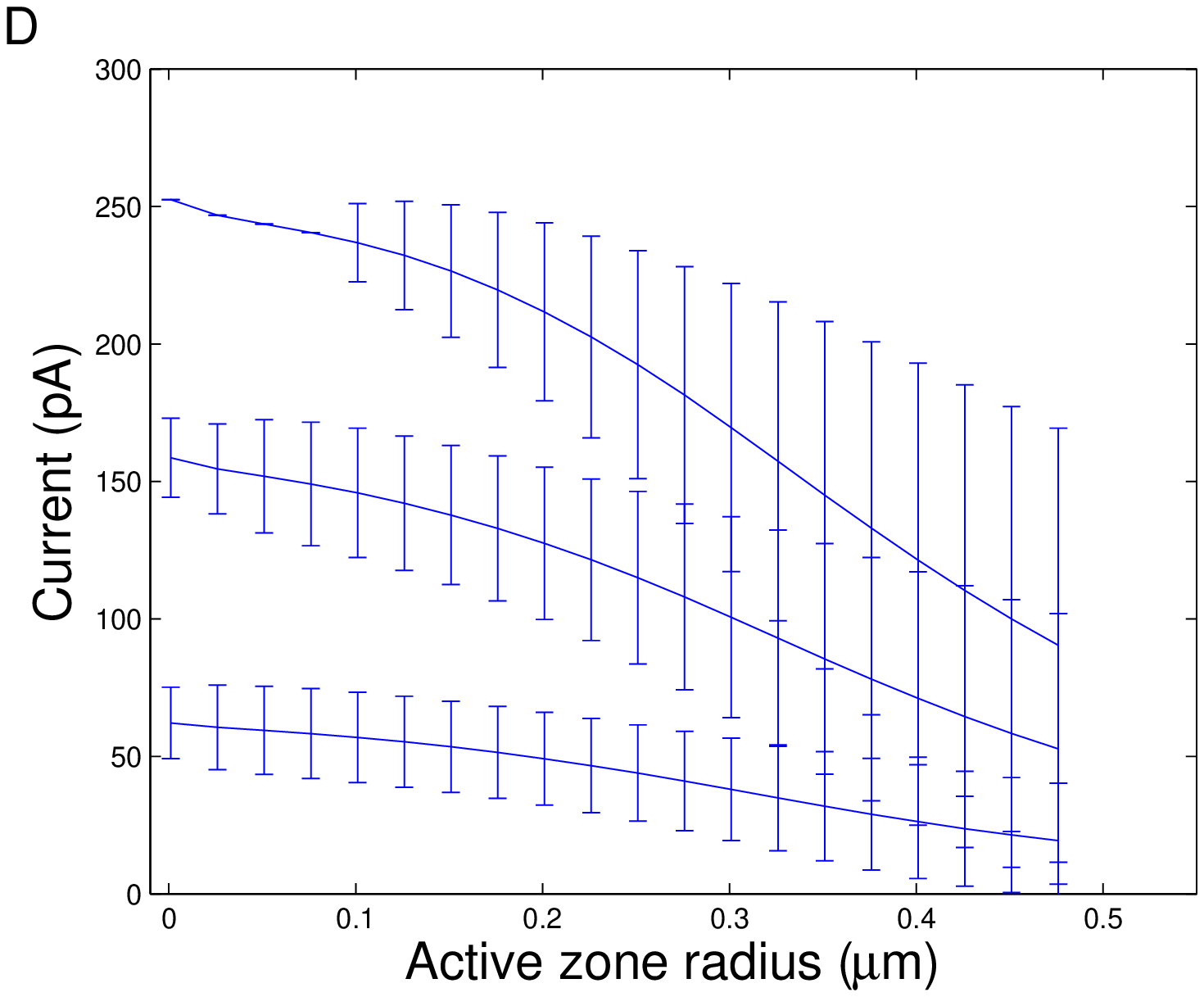,width=3.25in,clip=}\\
                    \epsfig{file=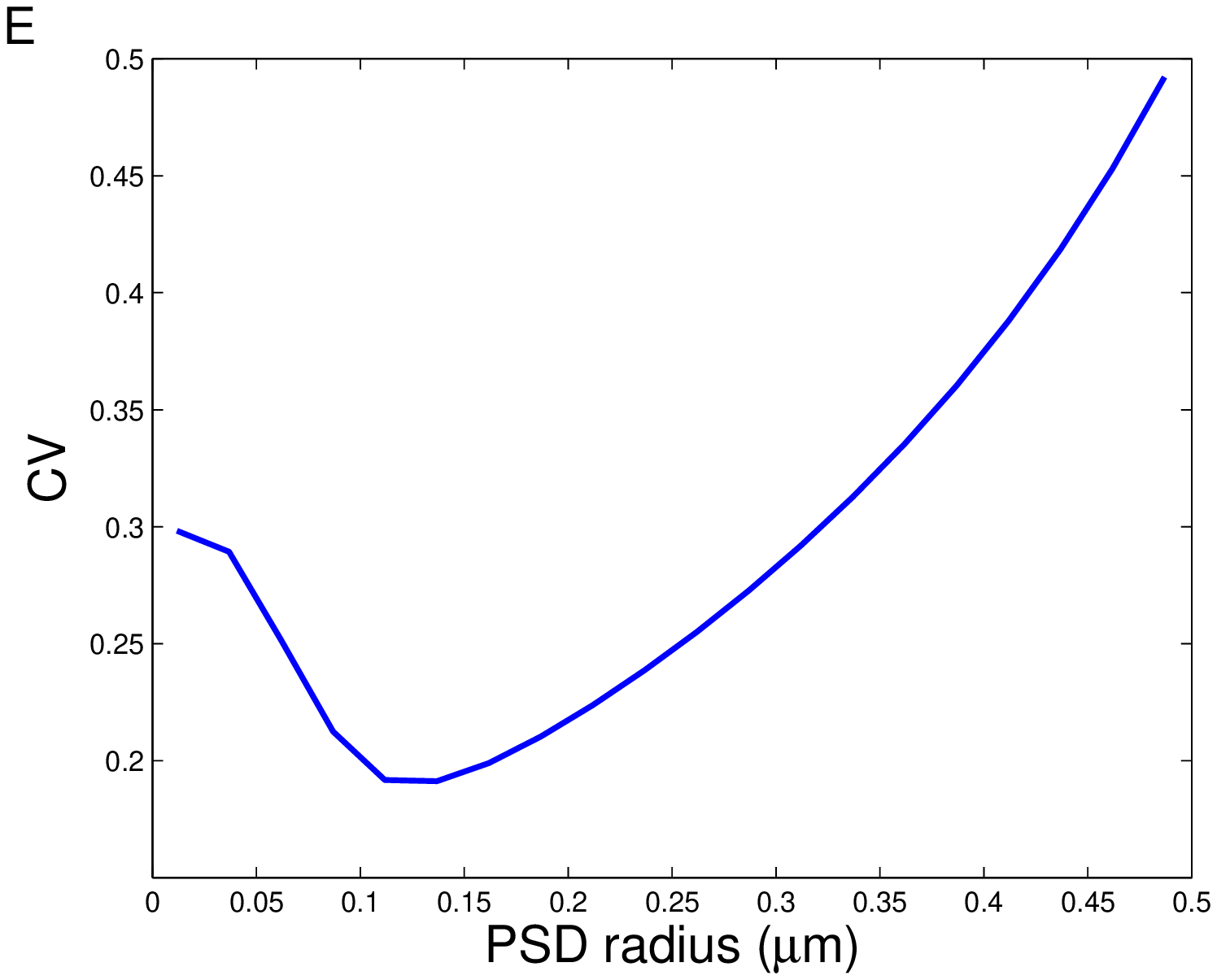,width=3.25in,clip=}&
                    \epsfig{file=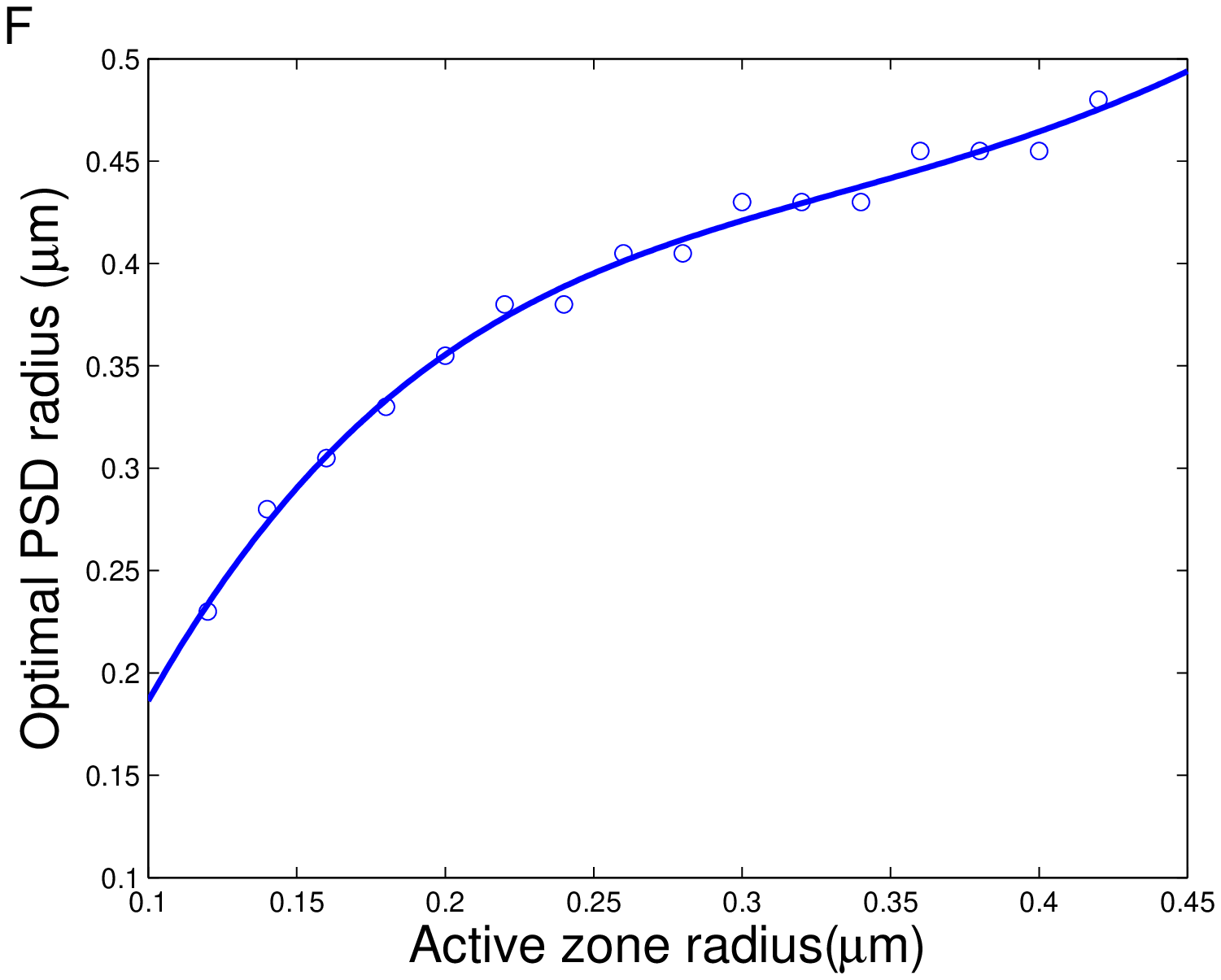,width=3.25in,clip=}
\end{tabular}
      \caption{}
      \label{figr3}
   \end{center}
\end{figure}

Since our analysis allows us to determine the relative influence of
the AZ and the PSD, we vary their respective sizes and we estimate
the consequences on the synaptic current. We first plotted in figure
\ref{figr3}A, the mean and variance of $I_s$ as a function of the
PSD radius when one vesicle is released at the center. We find that
for a given density of receptors,  the current is a decreasing
function of the PSD radius. In figure
\ref{figr3}B-C, we plotted the number of AMPARs bound
to two, three and four glutamate molecules when we fixes the AZ
radius to 50nm (blue) and 150nm (red). For a small AZ radius, the
AMPARs are in majority bound to four glutamate molecules and thus
the synaptic current amplitude is much higher compared to the case
of a large AZ radius. In that latter case, the current is primarily
generated by receptors bound to two glutamate molecules. We can now
use our refined analysis to re-interpret the large current
differences observed in figure \ref{figr3}A with the radii 50nm and
150nm.  This difference is due to the nonlinear properties of having
different conductivities, generated by the amount of bound glutamate
molecules. But in all cases, the synaptic current is a decreasing
function of AZ radius (Fig. \ref{figr3}D).

To asses the reliability of the synaptic response, we use the
coefficient of variation (CV=standard deviation over the mean) of
the synaptic  $I_s$. For a fixed AZ radius, we now show that the CV
has a minimum as function of the PSD radius: Indeed, using our
simulation results \ref{figr3}E, we find for example with an AZ of
radius 100 nm, that the CV reaches its minimum for a PSD radius of
120 nm. In order to confirm this result, we performed Brownian
simulations associated with our semi-analytical model. This Brownian
simulation is described as followed: glutamate molecules are
released from the AZ of radius $250nm$ and diffuse in a synapse of
radius $500nm$. We run simulations (fig.\ref{fig:simul}A) for
various PSD sizes and computed the associated CV (for an average of
100 samples). In the simulations, a receptor is modeled as a partial
absorber and when it binds to four glutamate molecules, it becomes
totally reflecting. The current is computed using equation
\ref{current}. Finally, any glutamate molecule reaching the lateral
boundary of the synapse is permanently absorbed. Using these rules,
We find (fig.\ref{fig:simul}B) that our initial analysis is
confirmed by the Brownian simulations .

\begin{figure}
   \begin{center}
   \begin{tabular}{ll}
      \epsfig{file=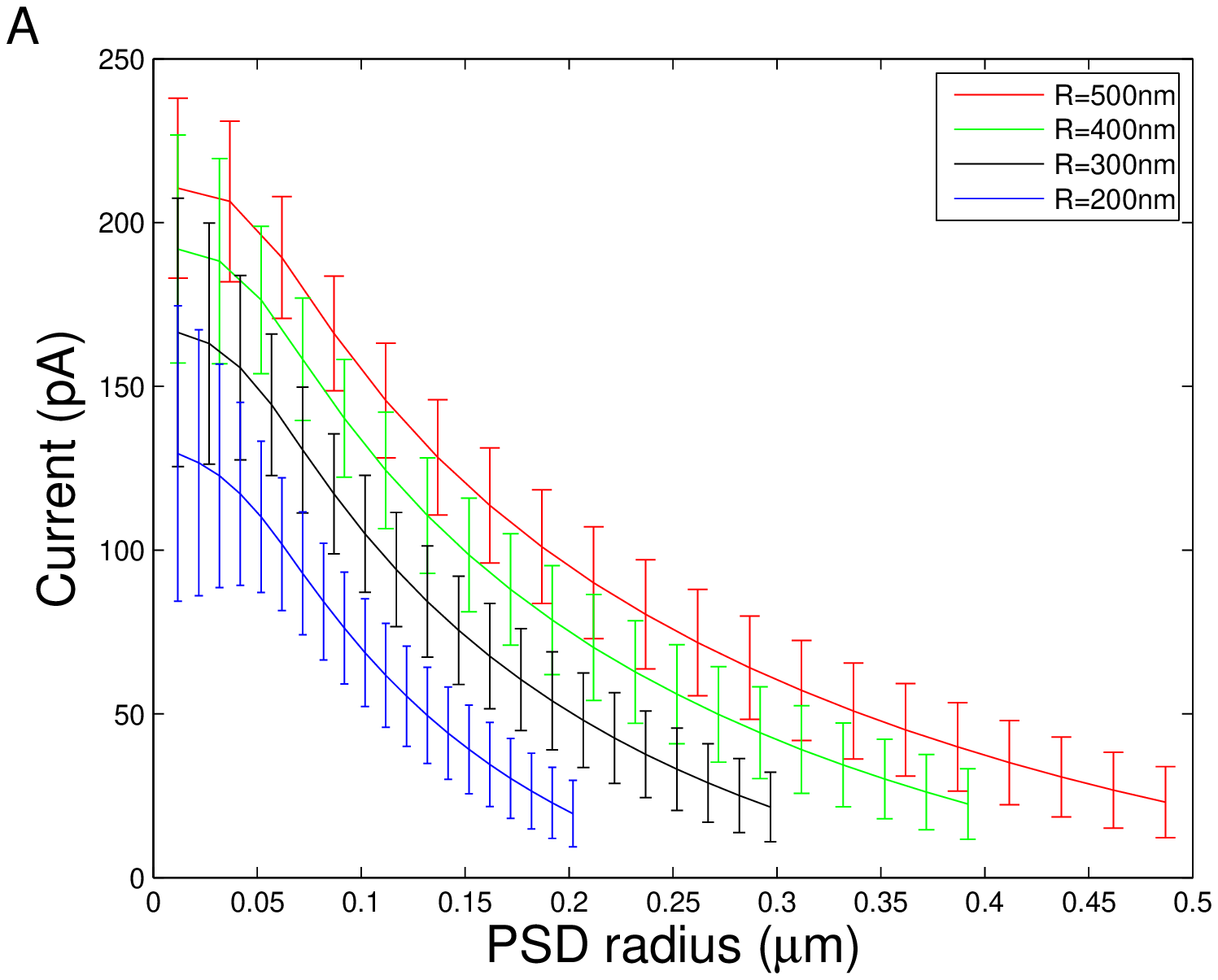,width=3.25in,clip=}&
        \epsfig{file=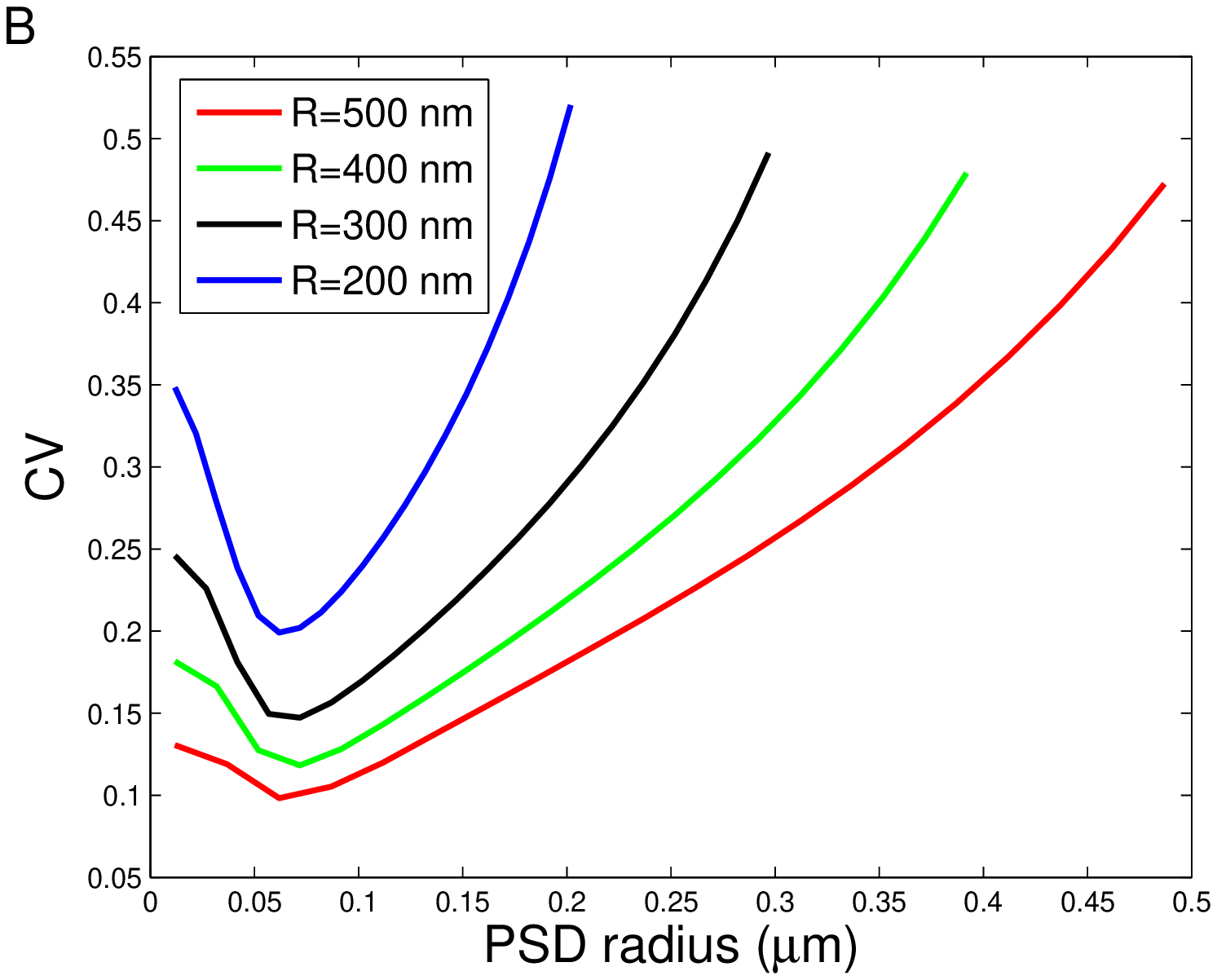,width=3.25in,clip=}\\
            \epsfig{file=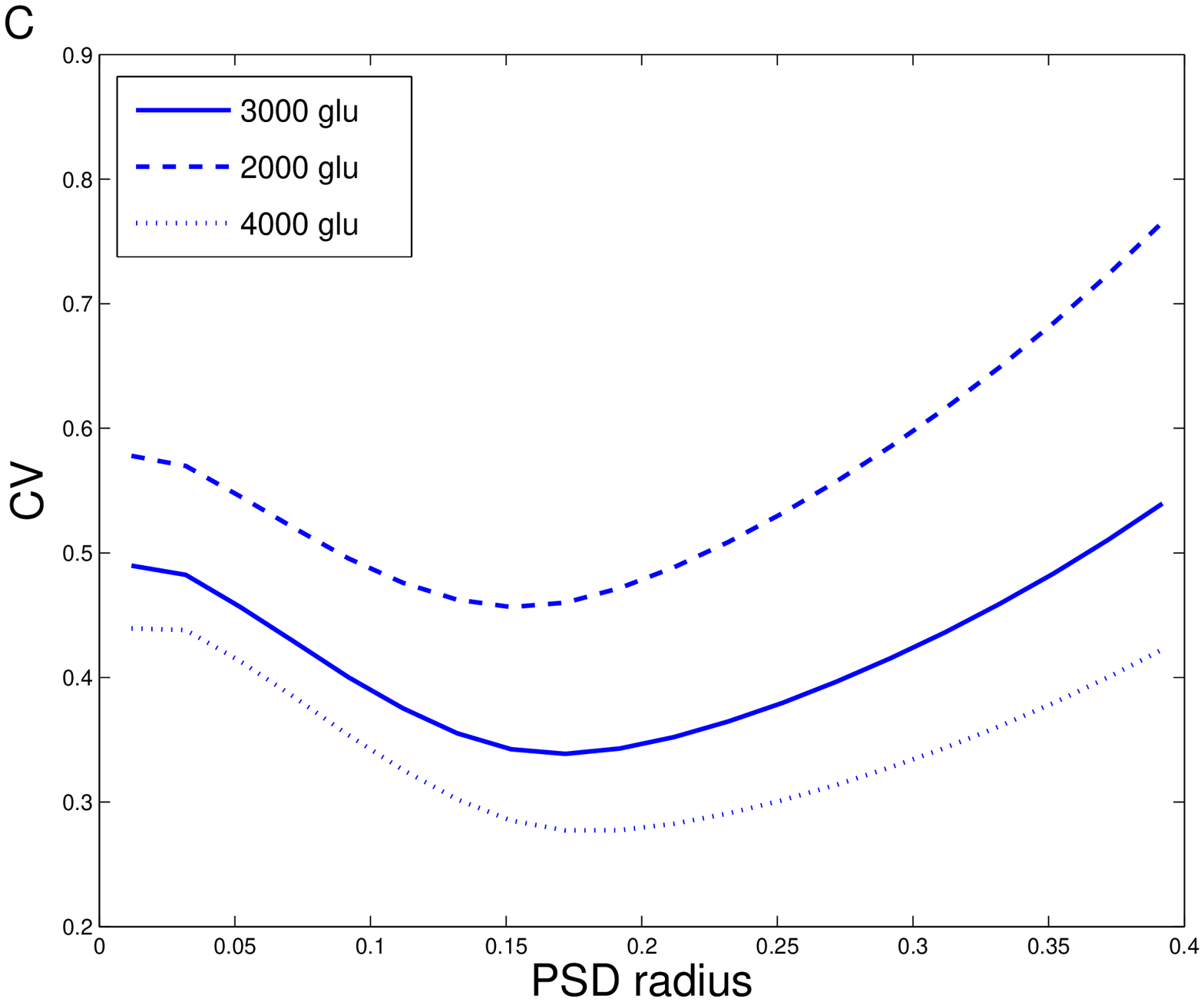,width=3.25in,clip=}&
                \epsfig{file=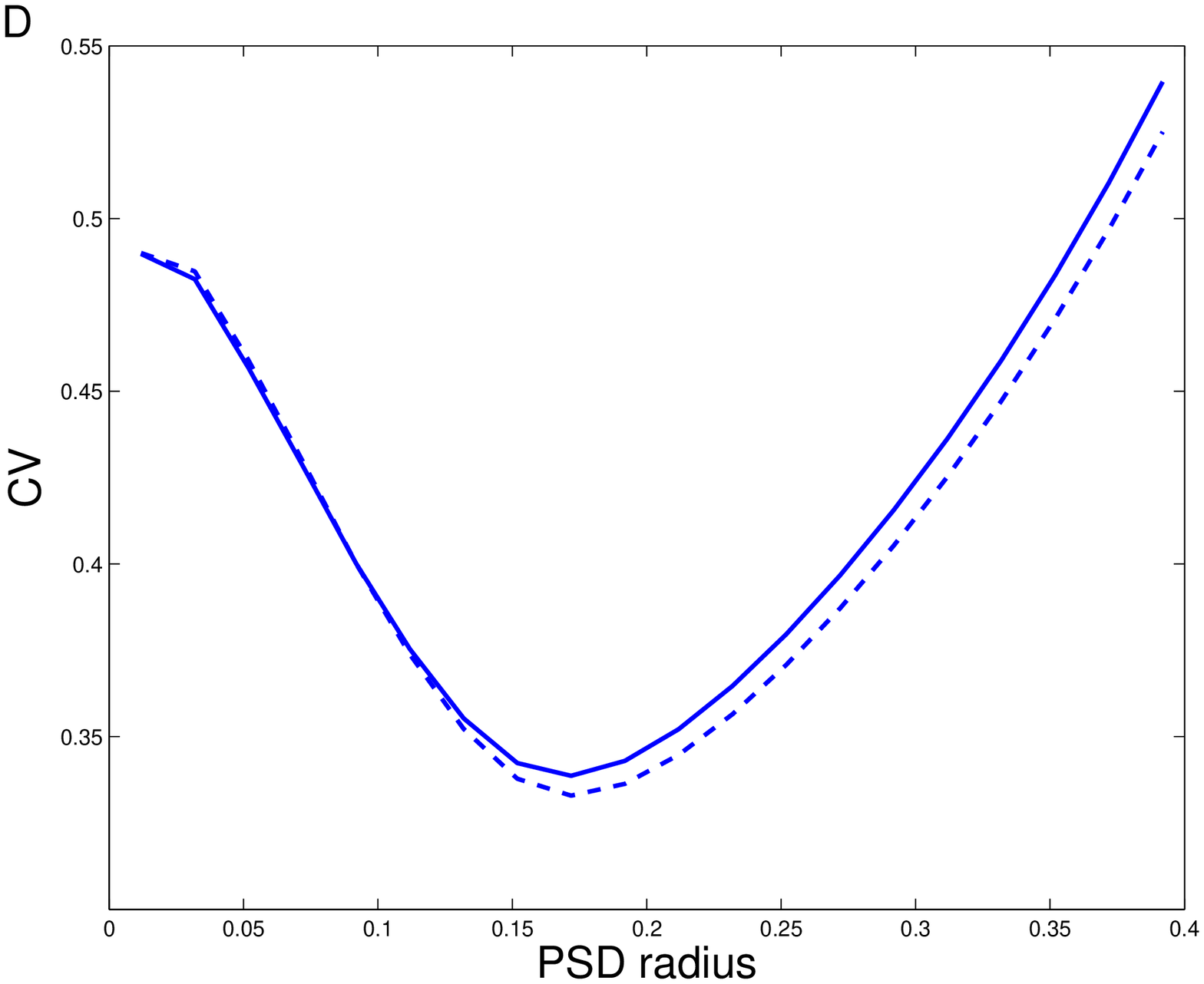,width=3.25in,clip=}\\
                    \end{tabular}
      \caption{}
      \label{figr4}
   \end{center}
\end{figure}

\begin{figure}
   \begin{center}
      \includegraphics*[width=3.25in]{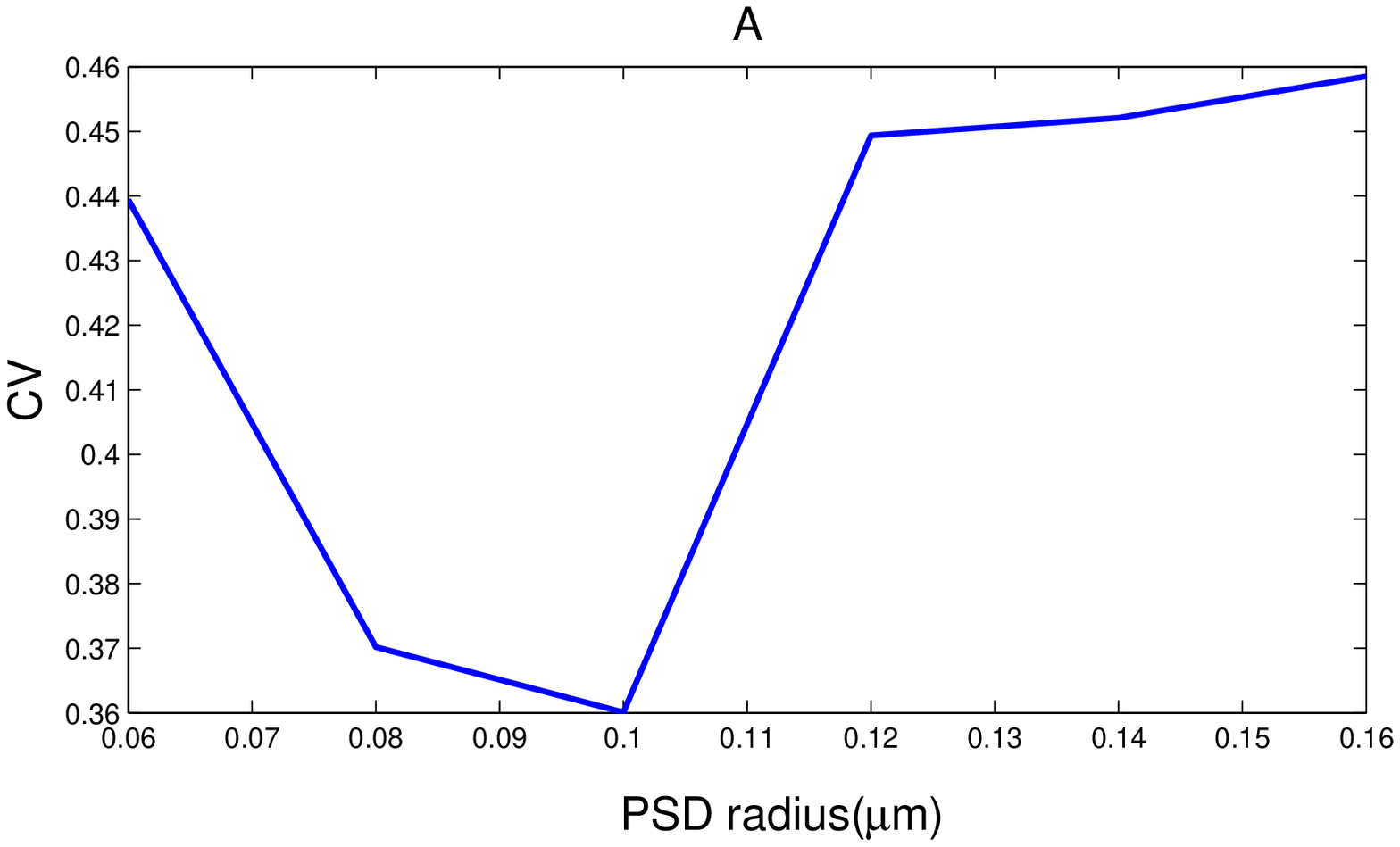}
        \includegraphics*[width=3.25in]{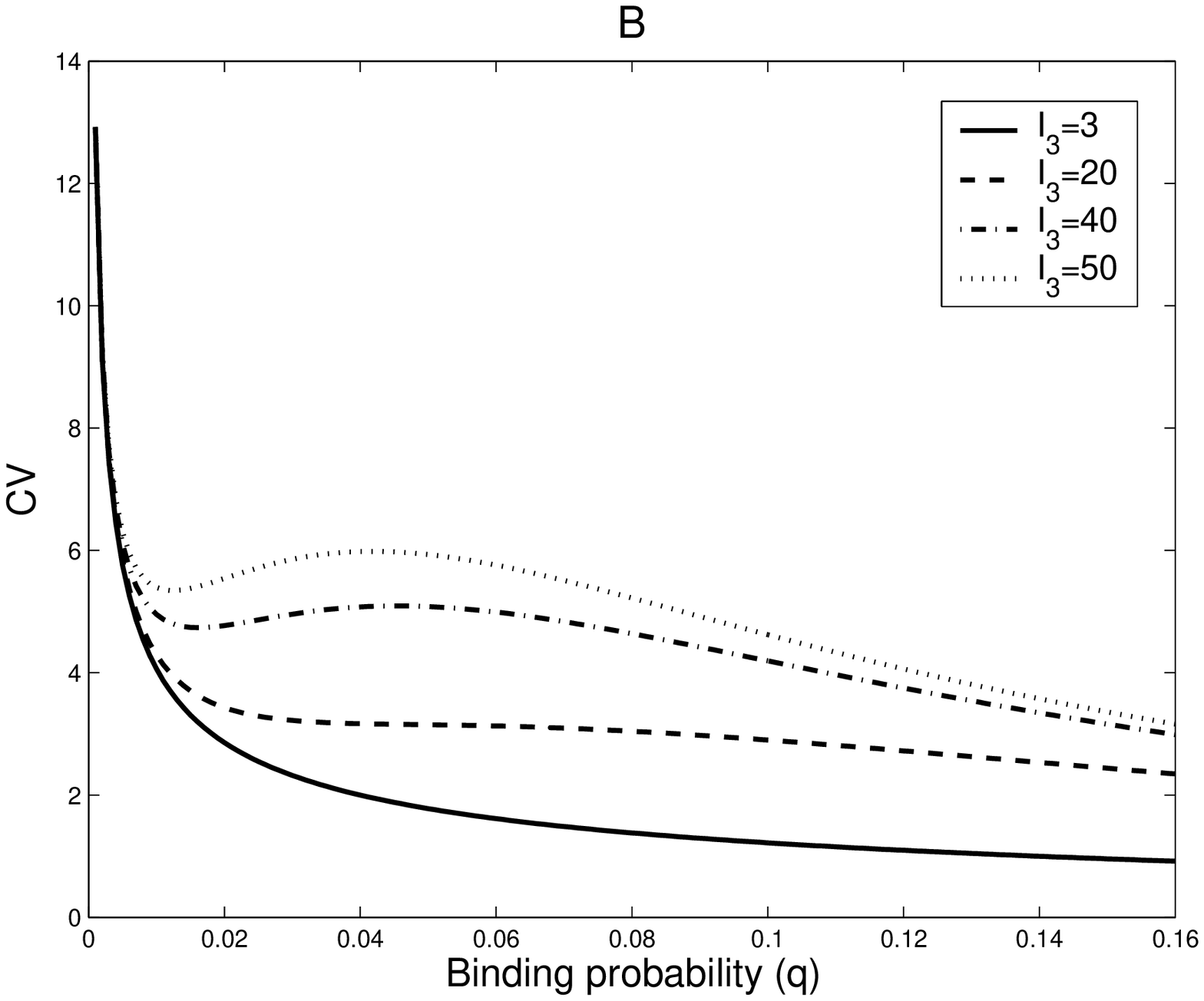}
      \caption{}
      \label{fig:simul}
   \end{center}
\end{figure}

To further investigate the relation between the radius of the PSD
and the AZ, we plotted (figure \ref{figr3}F) the optimal PSD radius
as a function of the AZ size. We find that the optimal PSD size
increases with the size of the AZ, but the relationship is a
nontrivial nonlinear. To see whether the size of a synapse affects
the CV curve, we fixed the AZ radius at 50 nm and plotted the
synaptic current as a function of the PSD radius for four different
synapses of sizes 200,300,400 and 500 nm (Fig.\ref{figr4}).
Interestingly, as shown in Fig.\ref{figr4}B, the value of the PSD
for which the optimal CV is achieved,  does not depend much on the
synaptic radius. We observe that the CV is a decreasing function of
the synaptic radius and thus large synapses are more reliable than
small ones. Finally, to assess the role of a possible fluctuation in
the number of glutamate molecules released from a vesicle, we show
in figure Fig.\ref{figr4}C the effect of three different
distributions for 2000, 3000 and 4000. This shows very little change
in the CV minimum phenomena. In Fig.\ref{figr4}D, we compare release
where the mean number is 3000 molecules with a variance of 500
Gaussian distributed (bold line) with a release of a fixed number
3000 (dashed line). We observe a small deviation, showing that in
this range of fluctuation, the number of released glutamate
molecules does not affect much the CV.

\section*{Discussion}
We have analyzed here the synaptic current $I_s$ starting from the
intrinsic biophysical properties of the AMPARs
\cite{Cull-candy2006}. We included the diffusion of glutamate
molecules in the narrow synaptic cleft and estimated the fraction
that binds AMPARs. Contrary to previous works
\cite{Sakmann,Barbour,Holmes,Franks}, our analysis does not use the
description of channels involving a time dependent multi-states
Markov chain, instead we use a multi-conductance approach mixed with
a time independent receptor description. In addition, we neglected
any possible interactions between bound receptor subunits that would
affect the probability for a free subunit to bind a glutamate
molecule. This approach allows us to obtain semi-analytical results
about the synaptic current and to account for the nonlinear effect
due to the multi-conductance states of bound AMPARs. We also studied
the role of the cleft geometry and explore the role of several
parameters.

For example, we have found here that modulating the PSD size can
affect the synaptic current amplitude: the current depends on the
relative size of the PSD to the AZ (Fig. \ref{figr3}A,D). As a
consequence, because the PSD could be reorganized, while the total
number of receptors remains constant, we propose that this direct
change in the PSD geometry can modulate the synaptic current. These
functional consequences were anticipated in recent experiments using
GFP-tagged PSD-95 \cite{Elhers1}. As suggested there, the PSD size
is in constant remodeling. Thus, combined with our analysis, we
propose that  synaptic changes are a source of synaptic current
modulation. It is not clear what are the reasons for these changes,
but they could be induced by long term potentiation (LTP) or
depression protocols \cite{Elhers1}. Changing PSD size, while
keeping AZ fixed can be seen as a form of plasticity induced by
structural remodeling without any change in the number of receptors.
These changes can, for example, be induced by actin dynamics, which
is correlated with spine shape changes
\cite{Cingolani,Okamoto}.

We have not accounted here neither for glial cells nor neuronal
transporters, that only weakly affect direct synaptic transmission
and the number of open AMPARs. The effect is of the order $10\%-
15\%$ (\cite{Barbour},\cite{Franks2} Figure 9,
\cite{Diamond97}, \cite{IsaacsonNicoll},\cite{TzingounisWadiche07}).
However, in some pathological conditions, related to a glial
reorganization, glial transporters can directly modulate synaptic
transmission \cite{parnash}, which changes drastically the present
results.

Another aspect of our modeling approach relies on the description of
AMPARs. Indeed, we computed the synaptic current from conductances
originating from patch-clamp experiments of isolated AMPARs
\cite{Cull-candy2006,Smith}, where the relation between the number
of bound glutamates and the associated conductances was obtained for
different fixed glutamate concentrations. Although more than four
conductance levels have been reported, it is still unclear how to
relate them to the number of bound glutamate molecules. In that
context, it would be interesting to design a specific experiment to
measure simultaneously on single AMPAR the number of bound glutamate
and the associated current. In addition, having four glutamate bound
to a single receptor lead to an amplitude of $13\,pA$, which has to
be compared with two bound AMPARs to two glutamates leading to
$2\cdot 4=8\,pA$). This difference suggests that binding four
glutamates to a single AMPAR has a nonlinear effect and is different
from having two receptors bound by two glutamates, as reported in
figures \ref{figr3}A-\ref{figr3}B and \ref{figr3}C. To show that the
minimum of the CV comes from this nonlinearity,  we shall now
consider a reduced model for single receptor with three states,
immersed in an ensemble of $N$ binding molecules. Depending on the
number of bounds, the receptor can be in one of the three states,
where the current $I$ can switch between the three values
$[I_1,I_2,I_3]$. Each molecule binds with a probability $q
\in [0,1]$ and the probability of $I$ is given by
\beq
\label{toymodelprob}
p(I=I_k)=\left\{\begin{array}{cc}\noverk{N}{k}q^k(1-q)^{N-k}& k=1,2\\
&\\1-p(I=I_1)-p(I=I_2)-(1-q)^N & k=3.\end{array}\right.
 \eeq
where the probability for one and two is given by the Binomial law
 while to compute the probability for the current $I_3$, we use that
it is the difference of the first twos minus the probability that no
molecules are bound. We computed analytically the CV of the current
$I$ as a function of the probability $q$ and we observed
(Fig.\ref{fig:simul}B) that only for some range of the parameters
such as $I_1,I_2\ll I_3$, the CV presented a local minimum. This
results show this simple model captures the features for a minimum.
In addition,  in that model changing the parameter $q$ is equivalent
to vary AZ or the PSD radius.

To conclude, it is still unclear what defines the detection
threshold of a post-synaptic neuron. Indeed, this current is
mediated by the number of open AMPARs and the associated
conductances. When an AMPAR is maintained at the PSD by scaffolding
molecules such as PSD-95, located just underneath the location of
vesiclar released (a signal that can be mediated by N-cadherin
molecules \cite{El-Husseini}), the probability of glutamate binding
is maximal and thus the AMPARs  will report more accurately this
vesicular event. We predict that this effect will be increased when
the scaffolding molecules PSD-95 will be over-expressed. Indeed the
over-expression  will results in an increasing of the number of
anchored receptors and their relative location in comparison with
the pre-synaptic terminal. In contrast, in PSD-95 knockdown (by
shRNA \cite{Elias}), the detection threshold was found to be lower,
due to a decrease in the number of AMPARs and a disperse AMPAR
configuration.

A synapse is an unreliable device but the variability is reduced
when the AZ and the PSD are apposed with a precise relationship
between their radius (Fig. \ref{figr3}F). However, if the vesicular
release is spread over the AZ, receptors over the PSD can detect a
vesicular event, but it might be very small. The regulation of the
PSD size should thus be a fundamental parameter comparable to
increasing the AMPAR number, which is the molecular basis for robust
synaptic plasticity. Finally, from a biophysical point of view, the
partial reflection constant that have introduced for the simulation
of the AMPARs reflects both the energy activation barrier of
glutamate-AMPAR interaction and the probability to find a receptor
inside the PSD. The concept of alignment of vesicular release domain
with the PSD was already suggested in \cite{Berger} without
quantitative analysis. Thus changing the PSD shape can modulate the
synaptic current and thus can be considered as a source of
plasticity. It would be interesting to analyze the molecular
mechanisms responsible for such changes
\cite{Elhers1}.

\section{Appendix}

\subsection*{Summary of our methodology approach}
We shall now summarize our methodology and the organization of this
appendix. Our new method consists in combining analytical
computations and numerical simulations to estimate the probability
that a combination of glutamate molecules activate an AMPAR, with
direct experimental measured conductances \cite{Cull-candy2006}. The
methodology to estimate the distribution of bound AMPARs is first to
compute the probability that glutamate molecule binds a receptor.
For this we solve in appendix \ref{appen1}, the probability equation
and calculate the flux of receptors to a given region (the PSD) and
the total flux through all the synaptic cleft as a function of
variable x, position where glutamate molecules are released. The
ratio of the fluxes is the probability. Second, because the height
of the synaptic cleft is small, the solution u of equation \ref{eq2}
is close to its average over the z-direction and we use this
property to perform our computations. Third, we consider that the
AMPARs are uniformly distributed over the PSD and we use a
homogenization procedure to replace by a partial reflecting boundary
condition summarized in the constant $\kappa$, a complicated
boundary condition, where sometimes a glutamate molecule would bind
to an AMPAR receptor with a given probability of activation and
sometimes it would be reflected when hitting parts of the PSD
containing no receptors. Fourth, the rational expression for
$\kappa$ is derived in appendix \ref{appen2} and the associated
mathematical derivation is given in appendix
\ref{appen4}. To check the validity of our computation, we compare
our results with Brownian simulation, described in appendix
\ref{comparisonBM}. Finally, we use published data to obtain an
approximate value for $\kappa$ and relate it to the activation and
the effective binding size of a single AMPAR (appendix
\ref{effective}).

\subsection{ Analytical expression for the binding probability of a glutamate molecule to the PSD, using an
averaging method.} \label{appen1}
We present here an averaging method to obtain an explicit expression
for the probability $p(\x_0)$ that a glutamate binds one of the AMPA
receptors before it escapes. To compute the total number of bound
glutamate molecules to the AMPARs, we solve the steady state
diffusion equation in the cylindrical synaptic cleft geometry, where
glutamate is released at $\x_0$. Using an averaging method, we
analyze equation

\beq \label{eq2}
 &&D\Delta u(r,z)=-\delta(r)\delta(z-z_0)\mbox{ for }
\{(r,z)|r\in[0,R),z\in(0,h)\} \\
& &\nonumber\\
&&\left.\frac{\partial u(r,z)}{\partial \nu}\right|_{\partial
\Omega_r} =0, \,
\left.u(r,z)\right|_{r=R}=0 \nonumber \\
 & & \nonumber \\
&&D\left.\frac{\partial u(r,z)}{\partial \nu}\right|_{r<L,z=0}
=-\kappa u(r,z),\nonumber
  \eeq
In cylindrical coordinates, the average
$\bar{u}(r)=\frac{1}{h}\int_0^h u(r,z)dz$ satisfies
\beq\label{app1:main eq}
D\left(\bar{u}''(r)+\frac{1}{r}\bar{u}'(r)-\frac{1}{h}\frac{\partial}{\partial
z}u(r,0)\chi_{[0,L]}(r)\right)=-\frac{1}{rh}\delta(r),
\eeq
Using the boundary conditions on $\partial \Omega_{PSD}$ for
${r<L,z=0}$, we express $\frac{\partial}{\partial
z}u(r,0))\chi_{[0,L]}(r)=0$ in terms of $\bar {u}$:
\beqq
u(r,z)\approx u(r,0)+\frac{\partial u(r,0)}{\partial z}z+O(h^2).
\eeqq
Integrating the Taylor expansion with respect to $z$, and using that
for $u(r,0)=\frac{D}{\kappa}\frac{\partial u(r,0)}{\partial z}$ and
$r<L$, \beq\label{app1:eqfor u_z} \bar {u}\approx\frac{\partial
u(r,0)}{\partial z}\left(\frac{D}{\kappa}+\frac{h}{2}\right),
\eeq
by substituting eq.(\ref{app1:eqfor u_z}) in (\ref{app1:main eq}),
we get
\beqq
D\left(\bar{u}''(r)+\frac{1}{r}\bar{u}'(r)-\frac{2\kappa}{h(2D+\kappa
h)}\bar{u}\chi_{[0,L]}(r)\right)=-\frac{1}{rh}\delta(r).
\eeqq
The solution is given by
\beqq
\bar{u}(r)=\left\{\begin{array}{cc}\frac{1}{2\pi D}K_0(\alpha
r)+AI_0(\alpha r) &0<r<L\\ &\\ \ds{C\log\frac r R} &L<r<R
\end{array}\right.,
\eeqq where \beqq \alpha=\sqrt{\frac{2\kappa}{h(2D+\kappa h)}}.
\eeqq
To determine the parameters $A$ and $C$, we use the continuity of
$\bar{u}$ and its derivative at $r=0$. We obtain the linear system
to invert:
\beqq
&&\frac{1}{D}K_0(\alpha L)+AI_0(\alpha
L) =C\log\frac L R\\
&& -\frac{1}{D}\alpha K_1(\alpha L)+A\alpha I_1(\alpha L)=C \frac
1 L.
\eeqq
The solution to these equation is given by
\beqq
&&\left(\begin{array}{c}A\\C\end{array}\right)=\left(\begin{array}{cc}\alpha
I_0(\alpha L)&-\log\frac L R\\\alpha I_1(\alpha L)&\frac 1
L\end{array}\right)^{-1}\left(\begin{array}{c}-\frac{1}{
D}K_0(\alpha L)\\\frac{1}{  D}\alpha K_1(\alpha
L)\end{array}\right)=\\&&\frac 1{D\alpha\left(\log\frac L R
I_1(\alpha L)-L^{-1}I_0(\alpha L)
\right)}\left(\begin{array}{c} {L^{-1}K_0(\alpha L)+\log\frac L R
K_1(\alpha L )}\\\alpha I_1(\alpha L)K_0(\alpha L)+\alpha I_0(\alpha
L)K_1(\alpha L )\end{array}\right)
\eeqq
To compute the probability $p(\x_0)$ we estimate two fluxes: first
at the PSD given by

\beqq 
 J_{PSD}=\kappa \int_{\partial\Omega_{PSD}}
udS=2\pi\kappa\int_0^L\frac{1}{ D}K_0(\alpha
r)rdr+2\pi\kappa\int_0^L AI_0(\alpha r)rdr \eeqq and second the
flux at the lateral boundary
\beqq
J_{Lat}= -2\pi D Rh
\bar{u}'(R)= -2\pi Dh C.
\eeqq
Thus the probability to hit one of the receptor, where the vesicle
is released at the center is given by
\beqq
p=\frac{J_{PSD}}{J_{PSD}+J_{Lat}}.
\eeqq

\subsubsection*{General location of vesicular release.}
To estimate the flux for a general location of vesicular release, we
use the general expression of the Laplacian  to rewrite
equation (\ref{eq2}),
\beq \label{app:eq_for_uncenter}\\
\frac{\partial^2\bar{u}(r,\theta)}{\partial
r^2}+\frac{1}{r}\frac{\partial\bar{u}(r,\theta)}{\partial
r}+\frac{1}{r^2}\frac{\partial^2\bar{u}(r,\theta)}{\partial^2
\theta}-\alpha^2\bar{u}(r,\theta)\chi_{[0,L]}=-\frac{1}{rhD}\delta(r-r_0)\delta(\theta-\theta_0)\nonumber
\eeq
We chose the release point on the line $\theta_0=0$, the solution
can be developed in a cosine series
\beq\label{app:expansion}
\bar{u}=\frac{a_0}{2}+\sum_{n=1}^\infty a_n\cos(n\theta).
\eeq
To estimate the flux,  we shall compute
\beqq
\kappa\int_0^{2\pi}\int_0^L\bar{u}(r,\theta)rdr\,
d\theta=\kappa\int_0^{2\pi}\int_0^L\frac{a_0(r,
r_0)}{2}rdr\,d\theta \eeqq By substituting the expansion
(\ref{app:expansion}) in equation(\ref{app:eq_for_uncenter}) and
integrating with respect to $\theta$ we found that $a_0$ satisfies
\beq\label{app:eqfora_0} \frac{d^2 a_0(r|r_0 )}{d
r^2}+\frac{1}{r}\frac{d a_0(r|r_0)}{d r} -\alpha^2
a_0(r|r_0)\chi_{[0,L]}=\frac{1}{D\pi hr}\delta(r-r_0). \eeq We
look for a solution of the form \beqq
a_0(r)=\left\{\begin{array}{cc}A_1
I_0(\alpha r)&0<r\leq r_0\\ &\\
A_2I_0(\alpha r)+B_2K_0(\alpha r)&r_0<r\leq L\\&\\
A_3\log(r/R)&L<r<R\end{array}\right.
\eeqq
When $r_0<L$, the continuity at the point $\x_0=(r_0, \theta_0=0)$
leads to
\beq\label{app:eq1of4} A_1 I_0(\alpha r)-A_2I_0(\alpha
r_0)-B_2K_0(\alpha r_0)=0
\eeq
while integrating equation(\ref{app:eqfora_0}) over
$(r_0-\eps,r_0+\eps)$, we obtain the condition
\beqq
\int_{r_0-\eps}^{r_0+\eps}\frac{d}{dr}(r\frac {d
a_0(r|r_0)}{dr})dr-\int_{r_0-\eps}^{r_0+\eps}\alpha^2
a_0(r|r_0)rdr=\frac{1}{D\pi h}
\eeqq
Taking the limit $\eps\to 0$, we get that
 \beq\label{app:eq2of4} A_2I_1(\alpha
r_0)-B_2K_1(\alpha r_0)-A_1I_1(\alpha r_0)=\frac{1}{\alpha D\pi h
r_0} \eeq The final condition comes from the interface $r=L$,
where we require the continuity of $a_0$ and its derivative with
respect to $r$. We get \beqq A_2I_0(\alpha L)+B_2K_0(\alpha
L)-A_3\log(L/R)=0\label{app:eq3of4}\\
A_2I_1(\alpha L)-B_2K_1(\alpha L)-A_3\frac{1}{\alpha
L}=0\label{app:eq4of4} \eeqq From equations
(\ref{app:eq1of4}-\ref{app:eq4of4}), we obtain four independent
equation for the coefficients $A_1,A_2,B_2,A_3$ and the net flux
can be express as
\begin{small}
\beq &&J_{PSD}(\x_0)=\kappa \int_{\partial\Omega_{PSD}}
u(r|\x_0)dS\label{JPSD1}  \\
&&=\kappa  \int_0^{r_0}(I_0(\alpha
r),0,0,0)\left(\begin{array}{cccc} I_0(\alpha r)&- I_0(\alpha
r_0)& -K_0(\alpha r_0)&0\\-I_1(\alpha r_0)& I_1(\alpha
r_0)&-K_1(\alpha r_0)&0\\0&I_0(\alpha L)&K_0(\alpha L)&
-\log(L/R)\\0&I_1(\alpha L)&-K_1(\alpha L)&-\frac{1}{\alpha
L}\end{array}\right)^{-1}\left(\begin{array}{c}0\\\frac{1}{\alpha
D}\\0\\0\end{array}\right)rdr\nonumber
\\[6pt]&&+\kappa\int_{r_0}^{L}(0,I_0(\alpha
r),K_0(\alpha r),0)\left(\begin{array}{cccc} I_0(\alpha r)&
-I_0(\alpha r_0)& -K_0(\alpha r_0)&0\\-I_1(\alpha r_0)& I_1(\alpha
r_0)&-K_1(\alpha r_0)&0\\0&I_0(\alpha L)&K_0(\alpha
L)&-\log(L/R)\\0&I_1(\alpha L)&-K_1(\alpha L)&-\frac{1}{\alpha
L}\end{array}\right)^{-1}\left(\begin{array}{c}0\\\frac{1}{\alpha
D}\\0\\0\end{array}\right)rdr.\nonumber
\eeq
\end{small}
 In
addition, the lateral flux is given by \beq\label{JLAT}
J_{Lat}(\x_0)= -2\pi D Rh \bar{u}'(R)= -2\pi D h A_3.
\eeq
Thus the total probability to be activated one receptor when the
neurotransmitters are released at position $\x_0$ is given by
\beq\label{proba3}
p(\x_0)=\frac{J_{PSD}(\x_0)}{J_{PSD}(\x_0)+J_{Lat}(\x_0)}. \eeq In
practice, we solve (\ref{app:eq1of4}-\ref{app:eq4of4})
numerically.

When a vesicle is released at a position outside the PSD,( $r_0\geq
L$), $a_0$ has the form
 \beqq
 a_0(r)=\left\{\begin{array}{cc}A_1 I_0(\alpha r)&0<L\leq
r_0\\&\\ A_2\log(r)+B_2&L<r \leq r_0\\&\\
A_3\log(r/R)&r_0<r<R.\end{array}\right. \eeqq Using similar
considerations as previously, the continuity conditions lead to
the set of equations: \beq
\label{linearsystem}\left\{\begin{array}{cccc} A_1 I_0(\alpha
L)-A_2\log(L)-B2&=&0\\
A_1 \alpha I_1(\alpha L)-\frac{A_2}{L}&=&0\\
A_2\log(r_0)+B_2-A_3\log\frac{r_0}{R}&=&0\\
-\frac{A_2}{r_0}+\frac{A_3}{r_0}&=&\frac{-1}{\pi D hr_0}
\end{array}\right.
\eeq Solving the linear system (\ref{linearsystem}), we obtain the
following expression for the flux:
\beq&&\\\label{JPSD2}
&&J_{PSD}(\x_0)=\int_0^L(I_0(\alpha
r),0,0,0)\left(\begin{array}{cccc}I_0(\alpha L)&-\log(L)&-1&0\\
\alpha I_1(\alpha L)&\frac{-1}{L}&0&0\\
0&\log(r_0)&1&-\log\frac{r_0}{R}\\0&-\frac{1}{r_0}&0&\frac{1}{r_0}
\end{array}\right)^{-1}\left(\begin{array}{c}
0\\0\\0\\\frac{-1}{ D hr_0}
\end{array}\right)rdr\nonumber
\eeq

\subsection{A partial absorbing boundary condition at the PSD}\label{appen2}
We present here our methodology to compute the partial absorbing
constant $\kappa$ for an ensemble of $N$ partially reflecting
receptors of size $a$ located on the PSD. When a glutamate molecule
hits a single receptor, it can sometimes be activated or not. This
condition at a single receptor is given by a partial absorbing
condition
\beqq
-D\frac{\partial p}{\partial n} =
\kappa_a p
\eeqq
where $\kappa_{a}$ is the AMPA partially-reflecting activation
barrier ($\kappa_{a}=0$ if there is no activation barrier and the
receptor is activated upon a glutamate hitting while for
$\kappa_{a}=\infty$, the barrier would be so large that every
glutamate molecule would only be reflected). The value of
$\kappa_{a}$ depends on the intrinsic properties of the AMPA binding
site.

To compute $\kappa$ the homogenized partial absorption
coefficient, we consider that all the receptors are located in the
PSD disk of radius $R_{PSD}$. The general partial absorbing
boundary condition would be
\beq\label{pa} -D\frac{\partial
p}{\partial n} = \kappa ~ p \hbox{ on the PSD}.
\eeq
Our criteria would be that the flux through the $N_a$ individual
receptor and the flux through the partially absorbing PSD should be
equal. To compute the flux through the partial absorbing PSD, we
solve the Mean First Passage Time equation with the boundary
condition \eqref{pa} instead of an absorbing boundary condition:
\beq
D\Delta u&=&-1 \hbox{ on } \Omega\label{pa2}\\
\frac{\partial u}{\partial n}&=& 0 \hbox{ on } \partial \Omega_r\nonumber\\
-D\frac{\partial u}{\partial n} &=& \kappa~ u \hbox{ on the
PSD}\nonumber
\eeq
To solve this equation, we use the standard method involving the
Neumann-Green function (\cite{PNAS,partialrefl}). In three
dimensions, we find that mean first passage time to the PSD is
approximated by
\beq
\label{eetim} u(\x) \approx
\frac{|\Omega|}{D}\left(\frac{1}{4R_{PSD}}+\frac{D}{2\pi\kappa
R^2_{PSD}}\right)
\eeq
Thus the flux per particles is
\beqq
J=\frac{1}{u(\x)} \approx
\frac{D}{|\Omega|}\frac{1}{\frac{1}{4R_{PSD}}+\frac{D}{2\pi\kappa
R^2_{PSD}}}.
\eeqq
First let us consider the flux on $N_a$ AMPA receptors of size a in
a disk of size $R$, which are fully absorbing. Then, the MFPT is
\beq\label{eq45}
\tau =\frac{|\Omega|}{4R_{PSD}D}
\frac{N_aa+f(\sigma)R}{N_aa}
\eeq
Thus when we equal relation (\ref{eq45}) with (\ref{eetim}), we get
the relation
\beqq
\frac{|\Omega|}{D}\left(\frac{1}{4R_{PSD}}+\frac{D}{2\pi\kappa_{P}R^2_{PSD}}\right)=\frac{|\Omega|}{4R_{PSD}D}
\frac{N_aa+f(\sigma)R_{PSD}}{N_aa}
\eeqq
which leads to the expression for the partial homogenization
constant
\beqq\kappa =\frac{D}{2\pi R^2_{PSD}}\frac{N_aa}{f(\sigma)}
\eeqq
where $f(\sigma)=1-\sigma,$ $\sigma=N_aa^2/R^2_{PSD}.$ Now in
general, \cite{Zwanzig,Szabo,Dulko}, we obtain the following relation
\beqq
\tau
=\frac{|\Omega|}{D}\left(\frac{1}{4R_{PSD}}+\frac{f}{N_aa}+\frac{D}{\kappa_a
2\pi a^2 N_a}\right)
\eeqq
and thus \beq\label{formula kappa_P}
\kappa =\frac{D}{2\pi
R^2_{PSD}}\frac{1}{\frac{f(\sigma)}{N_aa}+\frac{D}{\kappa_a 2 \pi
a^2N_a} }
\eeq

\subsection{Comparison with Brownian simulations}
\label{comparisonBM}

Because our analytical analysis contains several approximations such
as the averaging over the cleft height or the coefficients
$A_1,A_2,B_2,A_3$ are approximated numerically, we evaluated the
accuracy of our analysis by comparing the probability (\ref{proba3})
with Brownian simulations (see figure
\ref{figureappendix}). We simulated Brownian particles in the same
cylindrical domain as the one used for the analytical computation
with an absorbing lateral boundary condition. We put at the PSD a
partial absorbing boundary with the condition $-D\frac{\partial
p}{\partial n}=-\kappa p$. At the particle level, we implemented the
reflection rule \cite{partialrefl}, in which particles hitting the
PSD boundary are reflected with a probability \beqq
P=\kappa\frac{\sqrt{\pi}}{\sqrt{D}} \eeqq and absorbed with the
complementary probability, where $D$ is the diffusion constant and
$\Delta t$ is the time step of the simulation. The scheme is
standard when the glutamate molecule is inside the cleft, but when
$x(t)+\sqrt{2D}\,\Delta  w<0$ at the PSD, then we use \beqq
 x(t+\Delta t)=\left\{ \begin{array}{l}
-(x(t)+\sqrt{2D}\,\Delta
 w)\quad\mbox{w.p.}\ 1-P\sqrt{\Delta t}\\
 \\
 \mbox{terminate trajectory otherwise}.
 \end{array}\right.
  \eeqq

\subsection{Estimation of the partial absorption rate $\kappa_a$ using experimental data.} \label{effective}
Using a Markovian kinetic model for the initial binding step of a
glutamate to an AMPAR, we estimate here the rate constant $\kappa_a$
and the homogenized coefficient $\kappa$ with the help equation
\ref{formula kappa_P}.

In a two state chain model, accounting for binding and unbinding of
a glutamate molecule to a receptor described as
\beq\label{Markovfig}
C\underset{k_{-1}}{\overset{k_1}{\rightleftarrows}}O,
\eeq
the forward binding rate $k_{1}$ is given in units of
$Molar~s^{-1}$. In one hand, the binding rate is calculated by the
flux formula as
\beq
J_{Markov}=k_1{A^{-1}N_g}V^{-1}\int_{\partial \Omega_a}p(x)dx\approx
k_1{A^{-1}N_g}V^{-1}\pi a^2 p(x),
\eeq
where $A$ is the Avogadro number, $p(x)$ is the density of glutamate
near the receptor and $V=\pi Rh^2$ is the volume of the synaptic
cleft. On the other hand, using the diffusion model, the flux term
is given
\beq
J_{diff}=N_g\kappa_a \int_{\partial \Omega_a}p(x)dx\approx
N_g\kappa_a\pi a^2 p(x),
\eeq
where $a$ is the radius of a receptor and $\partial\Omega_a$
represents the receptor surface. By equating the two fluxes
$J_{diff}$ and $J_{Markov}$, we obtain an expression for the partial
reflecting constant
\beqq \label{eq:k_amarkov}
\kappa_a={k_1 A^{-1}V^{-1}}.
\eeqq
Now, using the published value $k_1=10^7M^{-1}s^{-1}$ (taken from
\cite{Milstein}) we obtain that $\kappa_a\approx 1.06$. This
two-model chain state is a good enough approximation even in the
case where there are more states in the Markov chain. Indeed, the
transition rates to desensitization states are lower than the open
state so that in the short time scale, after binding (the time of
interest in our model), the probable state is the open state.

\subsection{Mathematical details for the computation of the partial rate
$\kappa$} \label{appen4}
We provide here the mathematical detail used in appendix
\label{appen4} for the expression of the partial absorbing constant
$\kappa$. We solve equation (\ref{pa2}) asymptotically using the
Green function:
\beqq
\Delta_{\x} N(\x,\y) &=& -\delta(\x-\y),
\quad\mbox{for}\quad \x,\y\in \Omega
 \\
&&\nonumber\\
\frac{\partial N(\x,\y)}{\partial \nu_{\x}} &=&
-\frac{1}{|\partial \Omega|},\quad\mbox{for} \quad \x \in \partial
\Omega,\ \y \in \Omega,
 \eeqq
If $\x$ or $\y$ (or both) are in $\partial \Omega$, then only a
half of any sufficiently small ball about a boundary point is
contained in $\Omega$, which means that the singularity of
Neumann's function is $\ds{ \frac1{2\pi|\x-\y|}}$. The Neumann's
function for $\y\in\partial \Omega$ is given by
\beqq
N(\x,\y) =
\frac{1}{2\pi |\x-\y|} + v(\x,\y),
\eeqq
where $v(\x,\y)$ is a regular function. Using Green's
identity and the boundary conditions (\ref{pa}), we obtain
 \beqq
u(\y) - \frac{1}{D} \int_{\Omega} N(\x ,\y) \,d\x  =
\int_{\partial \Omega} N(\x,\y) \frac{\partial u(\x )}{\partial
\nu}\,dS_{\x} +C,
 \eeqq
where
 \beqq
 C=\frac{1}{|\partial \Omega|}
\int_{
\partial \Omega} u(\x) \,d{\x}.
 \eeqq
The conservation of the flux leads (see \cite{SSH1}) to
\beq
\label{eq:comp} \int_{\partial \Omega_a} \frac{\partial
u(\x)}{\partial \nu_{\x}}\,dS_{\x} =- \frac{|\Omega|}{D},
\eeq
Following the argument in \cite{SSH1}, the function $N(\x,\y)$ is
integrable independent of $\partial \Omega_a$, whose integral is
uniformly bounded, whereas $C\to\infty$ as  $a\to0$. Setting
$g(\x)=\ds{\frac{\partial u(\x)}{\partial\nu_{\x}}}$ for $\x\in
\partial \Omega_a$ and using the boundary condition (\ref{pa}), we
obtain the integral equation for the flux density $g(\x)$ in
$\partial \Omega_a$,
 \beqq
 -\frac{D}{\kappa}g(\y) +\int_{\partial \Omega_a}N(\x,\y)g(\x)\,dS_{\x}=-C\quad\mbox{for}\quad\y\in
 \partial \Omega_a,
  \eeqq
Using the expansion of the flux
$g(\x)=g_0(\x)+g_1(\x)+g_2(\x)+\cdots$, where $g_{i+1}(\x)\ll
g_i(\x)$ for $a\to 0$ and choose
 \beqq
 g_0(\x)=\frac{-2\alpha}{a\pi \sqrt{1-\ds{\frac{|\x|^2}{a^2}}}}.
  \eeqq
It was shown in, \cite{partialrefl,SSH1} that if $\partial
\Omega_a$ is a circular disk of radius $a$, then
\beqq
\frac{1}{2\pi}\int_{\partial\Omega_a} \frac{g_0(\x )}{|\x -\y
|}\,dS_{\x} = \alpha\quad\mbox{for all}\quad\y\in\partial
\Omega_a.
\eeqq
Thus we approximate the solution by taking $\y$ at the origin, so
that  $g_0(0)=\frac{-2\alpha}{a\pi} $, thus we get
\beqq
C=\left(\frac{2D}{\kappa a \pi}+1\right) \alpha.
\eeqq
The flux condition (\ref{eq:comp}) gives
\beqq
\int_{\partial
\Omega_a} g_0(\x ) \,dS_{\x} = -4 a \alpha
\eeqq
We conclude that
\beqq u(\x) \approx C=|\Omega|\left(\frac{1}{2\pi\kappa
a^2}+\frac{1}{4Da}\right).
\eeqq


\clearpage
\section*{Figure Legends}
\subsubsection*{Figure~\ref{figr1}.}
{\bf{Schematic representation of the Synaptic cleft.}} The
synaptic cleft geometry is approximated as a narrow cylinder of
height $h$ and the PSD is positioned at the center of the
pre-synaptic terminal. We depicted a vesicle released at a
distance $r_0$ inside the Active Zone (AZ). Diffusing receptors
can either bind an AMPA receptor or diffuse away.\\

\clearpage
\subsubsection*{Figure~\ref{figr2}.}
{\bf The geometrical properties of the synaptic current.}
{\bf{A}}. We decompose the synaptic current $I_S$ in a sum of 2,3
and 4 glutamate bound glutamate molecules ($I_S=I_2+I_3+I_4$). In
the range of 3000-9000 glutamates, the contribution of each
configuration is $I_4>I_3>I_2$. The release is from the center.
{\bf B.} The synaptic current is plotted as a function of the
release distance from the center of the synapse for one, two and
three vesicles. In both graphs we used synapse of 500nm radius,
height 30nm, PSD radius of 300nm\\

\clearpage
\subsubsection*{Figure~\ref{fig:estimate_of_effectin ra}.}

Synaptic current (computed from \ref{syncurrent} ) versus the
number of glutamate molecules. We show current Vs. the number of
released glutamate molecules for different receptor radius,
$a=1.5nm,1.8nm$ and $2nm$ . For receptor effective binding radius
of $a=1.8nm$ saturation achieved at four vesicles.

 \clearpage
\subsubsection*{Figure~\ref{figr3}.}
{\bf {Optimal PSD radius}}. {\bf{A}}. The mean current and SD are
plotted as a function of the PSD size for three different active
zones (50 nm,100 nm and 150 nm) (the synaptic radius is 500 nm and
the height is 20 nm). {\bf{B, C}}. The mean number of AMPA
receptors bound by 2 (resp. 4) glutamate molecules as a function
of the PSD radius. {\bf{D}}. {\bf Current vs. Active Zone radius.}
The PSD size is fixed at 300 nm and each curve represents 1, 2 and
3 released vesicles. {\bf E. CV vs. PSD size.} The CV achieves a
minimum when the PSD and the active zone are approximately equal.
In that case the Active Zone is 100 nm and the CV minimum is
achieved for a PSD of radius of 120 nm. {F. \bf Optimal PSD
radius:} It is plotted as a function of the AZ radius obtained by
minimizing the CV for a fixed AZ.

\clearpage
\subsubsection*{Figure~\ref{figr4}.}
{\bf {Synaptic current for different synapse radius.}}.
 For a fixed active zone radius (50 nm), we plot
the synaptic current as a function of the PSD radius for four
different sizes of synapses 200,300,400 and 500 nm (the height is 30
nm). Doubling the size of the synapse leads to a current amplitude
that increases from 125 to around 190 pA for a small PSD radius.
{\bf{(B)}} The CV is plotted as a function of the PSD radius: the CV
minimum does not depend much on the synaptic radius. {\bf (C)} The
CV is plotted as a function of the PSD radius, for different number
of released glutamate molecules. There is an optimal PSD size, and
the CV decreases  as a function of the number of released glutamate
molecules. {\bf (D)} We compare the CV as a function of the PSD
radius curve when the number of released glutamate molecules is
distributed according to a Gaussian distribution with a mean 3000
and a standard deviation of 500 (bold line) with a fixed number of
3000 glutamate molecules (dashed line). Fluctuation in the number of
glutamate molecule in the vesicles does not affect much the general
behavior of the CV with respect to PSD size.

\clearpage
\subsubsection*{Figure~\ref{fig:simul}.}
{\bf {A. MC Simulation}}
 For a fixed active zone radius (300 nm) a
Monte-Carlo Brownian simulation of 500 molecules uniformly
released from the AZ.
  Each experiment has 100 repetitions at different PSD size.
  {\bf {B. CV analysis for a simple
  model}}. The model describes a random variable with three possible outcome values
 $I_1,I_2,I_3$. The probability function is given in (\ref{toymodelprob}).
The model describes a single receptor with three conductivity
levels, depends on the number of bound molecules, each can bind
the receptor with probability $q$. Each  curve represents the CV
Vs. probability of binding $q$ with different
 values of $I_3$, where $I_1=1,I_2=2$ are fixed. For large values of $I_3$, an optimal
 point emerges. The existence of optimal CV as function of binding probability
 is thus strongly connected to the
 nonlinear cooperative effect of multi-binding.

\begin{figure}
   \begin{center}
      \includegraphics*[width=3.25in]{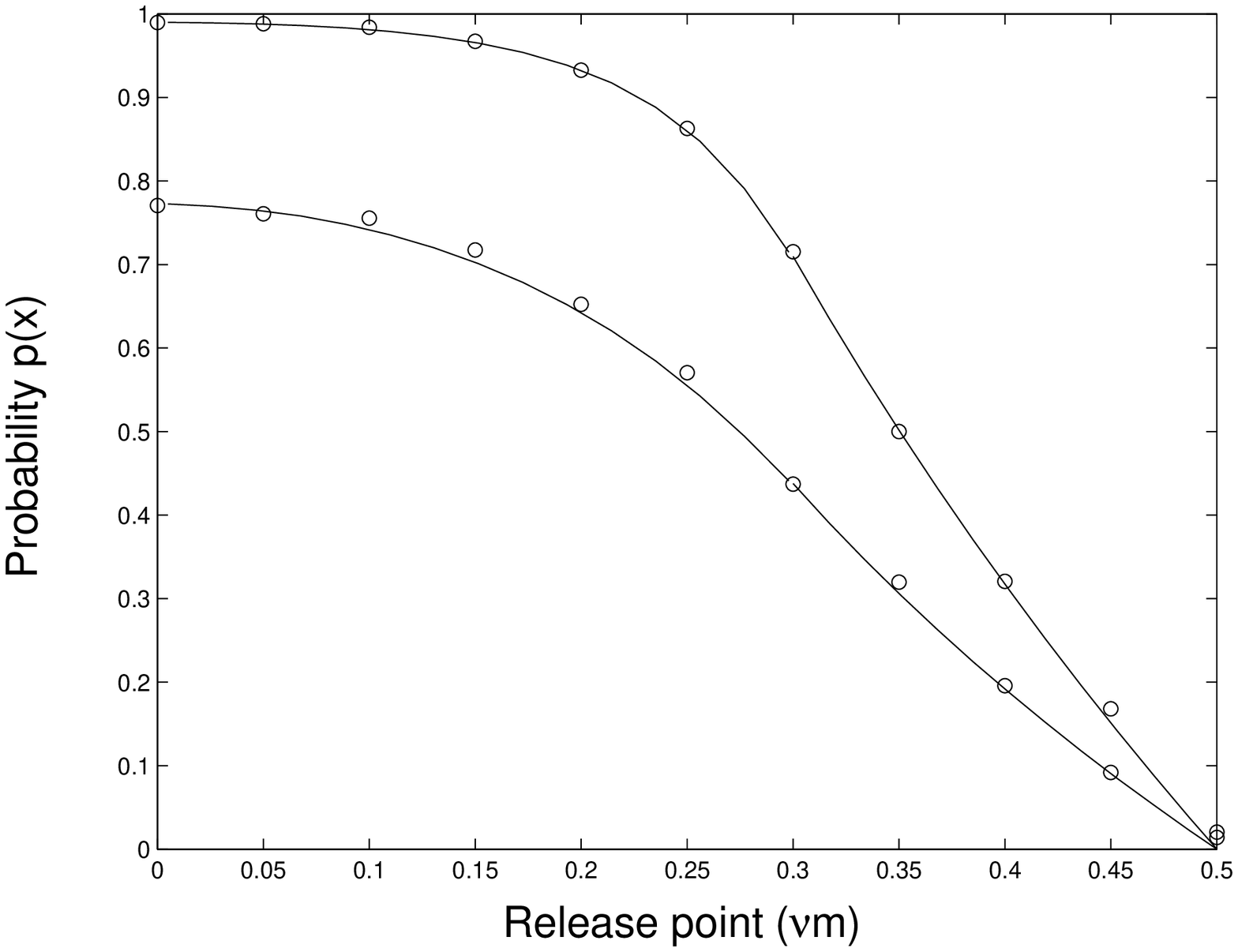}
      \caption{}
      \label{figureappendix}
   \end{center}
\end{figure}

\clearpage
\subsubsection*{Figure~\ref{figureappendix}.}
{\bf Probability $p(\x_0)$ to bind a receptor before exit the
cleft.} We compare here the analytic solution of (\ref{proba3})
(continuous line) with Brownian simulations (dots) in the
cylindrical domain of radius $R=0.5$, height $0.02$. The radius of
the PSD is $L=0.3$.
\clearpage
\clearpage

\clearpage
\clearpage
\clearpage
\clearpage

\clearpage

\end{document}